\documentclass[a4paper,10pt, reqno]{amsart}
\usepackage[a4paper, portrait]{geometry}	
		
\usepackage{amsfonts, amsmath, amssymb, amsthm}	
\usepackage{amsaddr}

\usepackage{bbm, dsfont, mathrsfs}		
\usepackage[T1]{fontenc}
\usepackage{url}

\usepackage{array, booktabs, multirow}
\usepackage[retainorgcmds]{IEEEtrantools}	
\usepackage{mathtools}
\usepackage{thmtools, thm-restate}

\usepackage{enumitem}
\usepackage{endnotes}				
	
\usepackage{footnote}

\usepackage{afterpage, pdflscape}
\usepackage[figuresright]{rotating}
\usepackage[usenames, svgnames, table]{xcolor}
	\definecolor{dark_gray}{gray}{0.05}
\usepackage{graphicx} 
	\usepackage{epstopdf}

\usepackage{abstract}

\usepackage{calc}			
\usepackage{xargs}			
\usepackage{setspace}		
\usepackage{comment}

\theoremstyle{plain}

\theoremstyle{definition}

\theoremstyle{remark}

\numberwithin{equation}{section}

\newcommand{\sqrts}[2][]{\,\sqrt[#1]{#2}\,}

\newcommand{\VaR}[2][]{\operatorname{VaR}_{#1}\!\left(#2\right)}

\def\E{\mathbb{E}}
\def\ead{\delta}
\def\el{l}
\def\given{\,|\,}

\def\lgd{\eta}

\def\P{\mathbb{P}}
\def\R{\mathbb{R}}

\usepackage[all, bottom]{background}
\SetBgContents{\parbox{\textwidth}{This article is \textcopyright{} Emerald Group Publishing and permission has been granted for this version to appear on \url{www.arxiv.org}. Emerald does not grant permission for this article to be further copied/distributed or hosted elsewhere without the express permission from Emerald Group Publishing Limited.\vspace{20pt}}}
\SetBgColor{dark_gray}
\SetBgScale{1.0}

\usepackage[authoryear, round]{natbib}

\begin{document}

\begin{titlepage}

\title[Assessing the Basel~II IRB Approach: Empirical Evidence from Australia]{}
\maketitle

\thispagestyle{empty}
\begin{center}

\vspace*{-40pt}
\textsc{\textbf{\Large Assessing the Basel~II Internal Ratings-Based Approach:\\Empirical Evidence from Australia}}
\par\vspace{20pt}
{\large Silvio Tarca\textsuperscript{~a,$\ast$} and Marek Rutkowski\textsuperscript{~a}}
\par\vspace{10pt}
{\small\textsuperscript{a~}\textit{School of Mathematics and Statistics F07, University of Sydney, NSW 2006, Australia.}}
\par\vspace{20pt}

\rule{\linewidth}{0.25mm}
\begin{abstract}
\vspace{4pt}\noindent\textbf{Purpose.} This study aims to render a fundamental assessment of the Basel~II internal ratings-based (IRB) approach by taking readings of the Australian banking sector since the implementation of Basel II and comparing them with signals from macroeconomic indicators, financial statistics and external credit ratings.  The IRB approach to capital adequacy for credit risk, which implements an asymptotic single risk factor (ASRF) model, plays an important role in protecting the Australian banking sector against insolvency.\\
\noindent\textbf{Methodology.} Realisations of the single systematic risk factor, interpreted as describing the prevailing state of the Australian economy, are recovered from the ASRF model and compared with macroeconomic indicators.  Similarly, estimates of distance-to-default, reflecting the capacity of the Australian banking sector to absorb credit losses, are recovered from the ASRF model and compared with financial statistics and external credit ratings.  With the implementation of Basel~II preceding the time when the effect of the financial crisis of 2007--09 was most acutely felt, we measure the impact of the crisis on the Australian banking sector.\\
\noindent\textbf{Findings.} Measurements from the ASRF model find general agreement with signals from macroeconomic indicators, financial statistics and external credit ratings.  This leads to a favourable assessment of the ASRF model for the purposes of capital allocation, performance attribution and risk monitoring.  Our empirical analysis reveals that the recent crisis imparted a mild stress on the Australian banking sector.\\
\noindent\textbf{Limitations.} Given the range of economic conditions, from mild contraction to moderate expansion, experienced in Australia since the implementation of Basel II, we cannot attest to the validity of the model specification of the IRB approach for its intended purpose of solvency assessment.\\
\noindent\textbf{Originality.} Access to internal bank data collected by the prudential regulator distinguishes our research from other empirical studies on the IRB approach and financial crisis of 2007--09.  We are not the first to attempt to measure the effects of the recent crisis, but we believe that we are the first to do so using regulatory data.

\vspace{8pt}\noindent \textsc{Keywords:} internal ratings-based (IRB) approach, asymptotic single risk factor (ASRF) model, credit value-at-risk (VaR), distance-to-default, reverse stress testing, financial crisis.
\end{abstract}
\rule{\linewidth}{0.25mm}

\vspace{20pt}
{\textnormal June 2016}
\end{center}

\vspace{50pt}
\noindent\rule{0.33\textwidth}{0.10mm}
\par{\footnotesize$^{\ast\,}$Corresponding author.  Telephone: +61 (0)8 8313 4178.  Email: \texttt{silvio.tarca@adelaide.edu.au}}

\end{titlepage}

\section{Introduction}\label{sect_intro}
Under the Basel~II Accord, authorised deposit-taking institutions (ADIs) determine regulatory capital for credit risk using either the standardised approach or, subject to approval, the internal ratings-based (IRB) approach.  The latter is more expensive to administer, but usually produces lower regulatory capital requirements than the former.  As a consequence, ADIs using the IRB approach may deploy their capital in pursuit of more (profitable) lending opportunities.  This paper examines the model specification of the IRB approach, the so-called asymptotic single risk factor (ASRF) model.  We render an assessment of the ASRF model by taking readings of the Australian banking sector since the implementation of Basel~II and comparing them with signals from macroeconomic indicators, financial statistics and external credit ratings.  The relatively short quarterly time series available leads to an intuitive assessment of the ASRF model rather than any formal testing of its efficacy.  A fundamental evaluation of the model specification of the IRB approach requires access to internal bank data, which are input to the ASRF model.  Support for this research by the Australian Prudential Regulation Authority (APRA) includes access to these data collected by APRA from the institutions that it supervises.

Upon implementation of Basel~II in the first quarter of 2008, APRA had granted the four largest Australian banks, designated ``major'' banks, approval to use the IRB approach to capital adequacy for credit risk. In order to take measurements from the ASRF model of the Australian banking sector, we aggregate data reported to APRA by the major banks on a quarterly basis since the implementation of Basel~II.  We contend that these data are representative of the Australian banking sector on the basis of the market dominance of the major banks, and the concentration of their regulatory capital assessed under the IRB approach.  The ASRF model calculates the expectation of credit losses conditional on a realisation of the single systematic risk factor, which is interpreted as describing the state of the economy, in order to assess regulatory capital charges.  Substituting credit losses incurred into the ASRF model, we solve for realisations of the single systematic risk factor describing the prevailing state of the Australian economy.  Then, substituting provisions and capital held against credit losses into the ASRF model, and translating realisations of the single systematic risk factor into distance-to-default, we measure the level of capitalisation of the Australian banking sector.  Our empirical findings comport with signals from macroeconomic indicators, financial statistics and external credit ratings.  While making observations about the prevailing state of Australia's economy and the solvency of its banking sector, we do not propose model enhancements, nor draw policy implications.  Although, our observations may open a debate on the model specification of the IRB approach or related regulatory policies.

The range of economic conditions, from mild contraction to moderate expansion, experienced in Australia since the first quarter of 2008 translate into observations away from the tail of the distribution of the single systematic risk factor, and hence the portfolio loss distribution.  Accordingly, we argue that our findings support a favourable assessment of the ASRF model for the purposes of capital allocation, performance attribution and risk monitoring --- management functions that need only be accurate in the measurement of relative risk under ``normal'' economic conditions.  The IRB approach, however, serves the sole purpose of solvency assessment, which requires precision in the measurement of absolute risk levels under stressed economic conditions, or tail events \citep{JF25}.  Therefore, we cannot attest to the suitability of the ASRF model for regulatory capital modelling.  Evaluating the model specification of the IRB approach for its intended purpose of solvency assessment would involve taking readings of banking jurisdictions that have experienced a full-blown economic or financial crisis since the implementation of Basel~II.  It would supplement the findings of this study.

The financial crisis of 2007--09, also known as the global financial crisis, precipitated the worst global recession since the Great Depression of the 1930s.  In response to the crisis, the so-called Basel~2.5 package and Basel~III Accord introduce reforms to address deficiencies in the Basel~II framework exposed by the recent crisis.  The Basel~2.5 reforms enhance minimum capital requirements, risk management practices and public disclosures in relation to risks arising from trading activities, securitisation and exposure to off-balance sheet vehicles.  The Basel~III reforms raise the quality and minimum required level of capital; promote the build up of capital buffers; establish a back-up minimum leverage ratio; improve liquidity and stabilise funding; and assess a regulatory capital surcharge on systemically important financial institutions.  These reforms complement, rather than supersede, Basel~II.  In particular, the ASRF model prescribed under the Basel~II IRB approach is unaltered by the introduction of Basel~2.5 and Basel~III.  It remains as relevant to solvency assessment today as it has been since its implementation in 2008.

From our fundamental assessment of the model specification of the IRB approach emerges a methodology for regulators to monitor the prevailing state of the economy as described by the single systematic risk factor, and the capacity of supervised banks to absorb credit losses as measured by distance-to-default.  Measurements from the ASRF model signalling an overheating economy and procyclical movements in capital bases, corroborated by macroeconomic performance indicators including rapidly accelerating credit growth, would prompt supervisors to lean against the prevailing winds by, for example, instructing banks to build up their countercyclical capital buffer introduced under Basel~III.

The effects of the financial crisis of 2007--09 were not felt evenly across the globe, and the Australian economy was largely spared.  With the implementation of Basel~II preceding the time when the effect of the crisis was most acutely felt, our empirical analysis reveals that the crisis imparted a mild stress on the Australian banking sector.  We are not the first to attempt to measure the effects of the recent crisis, but we believe that we are the first to do so using regulatory data.  In evaluating the model specification of the IRB approach we produce a fundamental assessment of the impact of the crisis on the Australian banking sector using internal bank data collected by APRA.  Other studies on the financial crisis rely on market data, macroeconomic indicators or published financial statistics.  In arguing the resilience of the Australian economy to the crisis, \citet{MM11}, and \citet{BD10} describe its performance primarily in terms of macroeconomic indicators and financial statistics.  Our measurements from the ASRF model of the Australian banking sector corroborate their observations.  \citet{AP12}, on the other hand, rely on market data and reach a markedly different conclusion about the condition of the Australian banking sector during the recent crisis.  We submit that their results are biased by plummeting market prices and spiking volatility reflecting the overreaction of market participants gripped by fear at the depths of the crisis.  

We begin in Section~\ref{sect_model_spec_irb} by outlining the model specification of the IRB approach, and discussing valid criticisms levelled against it.  In Section~\ref{sect_cap_adeq_rpt} we describe the Basel~II capital adequacy reporting of ADIs that supplies data for our empirical analysis.  Section~\ref{sect_perf_au_econ} contextualises the Australian economy over the past decade by comparing its performance with that of the United States and United Kingdom on macroeconomic indicators and financial statistics.  Explanations proffered for the recent performance of the Australian economy and its resilience to the financial crisis are discussed in some detail.  Measurements from the ASRF model of the Australian banking sector since the implementation of Basel~II are presented in Section~\ref{sect_asrf_au_banks}.  Realisations of the single systematic risk factor describing the prevailing state of the economy measure the impact of the crisis on the Australian banking sector, while estimates of distance-to-default reflect its capacity to absorb credit losses.  In a variation on the evaluation of distance-to-default, reverse stress testing explores stress events that would trigger material supervisory intervention.  We conclude by outlining the direction for future related research.

\section{Model Specification of the Internal Ratings-Based Approach}\label{sect_model_spec_irb}
The Basel~II IRB approach implements an asset value factor model of credit risk.  Asset value models posit that default or survival of a firm depends on the value of its assets at a given risk measurement horizon.  If the value of its assets is below a critical threshold, its default point, the firm defaults, otherwise it survives.  Asset value models have their roots in Merton's \citeyearpar{MRC74} seminal paper.  Factor models are a well established, computationally efficient technique for explaining dependence between variables \citep{BOW10}.

The model specification of the IRB approach assesses capital charges sufficient to absorb credit losses, and thus protect against insolvency, with a high level of confidence.  It applies value-at-risk (VaR), one of the most widely used measures in risk management, to assign a single numerical value to a random portfolio credit loss.  Let random variable $\mathcal{L}_{n}$ be the (dollar) loss on a credit portfolio comprising $n$ obligors over a given risk measurement horizon.   Adopting the convention that a loss is a positive number, we define \textit{credit VaR} at the confidence level $\alpha \in (0,1)$ over a given risk measurement horizon as the largest portfolio credit loss~$\el$ such that the probability of a loss~$\mathcal{L}_{n}$ exceeding~$\el$ is at most $(1\!-\!\alpha)$:
\begin{equation}\label{eqn_cred_var}
	\VaR[\alpha]{\mathcal{L}_{n}} = \inf\{\el \in \R\colon\P(\mathcal{L}_{n}>\el) \leq 1\!-\!\alpha\}.
\end{equation}
In probabilistic terms, $\VaR[\alpha]{\mathcal{L}_{n}}$ is simply the $\alpha$ quantile of the portfolio loss distribution, typically a high quantile that is rarely exceeded.  Although computationally expensive, Monte Carlo simulation is routinely employed to generate the empirical loss distribution and determine VaR of a credit portfolio.  An analytical model of the portfolio loss distribution, on the other hand, facilitates the fast calculation of credit VaR.  

The IRB approach rests on a proposition due to \citet{GMB03}, which leads to an analytical approximation to credit VaR, the so-called \textit{asymptotic single risk factor} (ASRF) model.  It assumes that: 
\begin{enumerate}[label=(\arabic*)]
 	\item\label{item_asymp_port}  Portfolios are infinitely fine-grained so that idiosyncratic risk is fully diversified away.
	\item\label{item_single_risk_factor}  A single systematic risk factor explains dependence across obligors.
\end{enumerate}
As a practical matter, credit portfolios of large, internationally active banks are typically near the asymptotic granularity of Condition~\ref{item_asymp_port}.  Furthermore, \citet{GL07} proposed a granularity adjustment for quantifying the contribution of name concentrations to portfolio risk, and hence assessing a capital charge for undiversified idiosyncratic risk.  So, moderate departures from asymptotic granularity need not pose an impediment to assessing ratings-based capital charges.  Condition~\ref{item_single_risk_factor}, on the other hand, is more stringent.  Denote by $Y$ the single factor common to all obligors, which may be interpreted as an underlying risk driver or economic factor, with each realisation describing a scenario of the economy.  \citeauthor{GMB03} established that the $\alpha$ quantile of the distribution of conditional expectation of portfolio credit loss may be substituted for the $\alpha$ quantile of the portfolio loss distribution:
\begin{equation}\label{eqn_var_cond_exp}
	\lim_{n\rightarrow\infty} \big|\VaR[\alpha]{\mathcal{L}_{n}} - \E[\mathcal{L}_{n} \given Y = \Phi^{-1}(1\!-\!\alpha)]\big| = 0.
\end{equation}

Suppose that \textit{exposure at default} (EAD) and \textit{loss given default} (LGD) are deterministic quantities, and denote by ${\ead_{i} \in \R_{+}}$ and ${\lgd_{i} \in [0,1]}$ the EAD and LGD, respectively, assigned to obligor~$i$.  Assume that latent random variables modelling the variability in obligors' asset values are standard Gaussian and conditionally independent given systematic risk factor~$Y$.  We denote by $\Phi$ the standard Gaussian distribution function.  Also, let ${\rho_{1}, \ldots, \rho_{n} \in (0,1)}$ be correlation parameters calibrated to market data where, for Gaussian processes, the pairwise correlation between obligors' asset values is equal to $\sqrts{\rho_{i}\rho_{j}}$.    Then, the \textit{conditional expectation of portfolio credit loss} is given by
\begin{equation}\label{eqn_cond_exp_port_loss}
	\E[\mathcal{L}_{n} \given Y=y] = \sum_{i = 1}^{n} \ead_{i}\lgd_{i}p_{i}(y) = \sum_{i = 1}^{n} \ead_{i}\lgd_{i}\Phi\left(\frac{\Phi^{-1}(p_{i}) - \sqrts{\rho_{i}}y}{\sqrts{1-\rho_{i}}}\right),
\end{equation}
where function $p_{i}(y)$ transforms $p_{i}$, the \textit{unconditional probability of default} of obligor~$i$, into the probability of default (PD) conditional on realisation~$y\in\R$ of systematic risk factor~$Y$, or \textit{conditional probability of default} of obligor~$i$.  This function, derived by \citet{VO02}, is the kernel of the model specification of the IRB approach.  We remark that conditional expectation function $\E[\mathcal{L}_{n} \given Y]$ is strictly decreasing in~$y$ --- conditional expectation of portfolio credit loss falls (respectively, rises) as the economy improves (deteriorates).  Thus, the $\alpha$ quantile of the distribution of $\E[\mathcal{L}_{n} \given Y]$ is associated with the $(1\!-\!\alpha)$ quantile of the distribution of $Y$.

\textit{Expected portfolio credit loss} is the average loss over all realisations~$y\in\R$ of systematic risk factor~$Y$, defined as
\begin{equation}\label{eqn_exp_port_loss}
	\E[\mathcal{L}_{n}] = \sum_{i = 1}^{n} \ead_{i}\lgd_{i}p_{i}.
\end{equation}

In keeping with the Basel~II Accord \citep{BCBS128}, we define \textit{unexpected loss} on a credit portfolio as the difference between $\VaR[\alpha]{\mathcal{L}_{n}}$ and $\E[\mathcal{L}_{n}]$.   Furthermore, we assume that ADIs set aside provisions for absorbing expected losses, and hold capital against unexpected losses.  This latter assumption is consistent with the prudential standards of APRA \citeyearpar{APS113}.  Appealing to Gordy's proposition, we substitute ${\E\big[L_{n} \given Y=\Phi^{-1}(1\!-\!\alpha)\big]}$ for $\VaR[\alpha]{L_{n}}$.  Hence, at the $\alpha$ confidence level over a given risk measurement horizon, capital held against unexpected credit losses on a portfolio comprising $n$ obligors is calculated as
\begin{IEEEeqnarray*}{rCl}
	 \mathcal{K}_{\alpha}(\mathcal{L}_{n}) & = & \E\big[\mathcal{L}_{n} \given Y=\Phi^{-1}(1\!-\!\alpha)\big] - \E[\mathcal{L}_{n}] \\
		& = & \sum_{i = 1}^{n} \ead_{i}\lgd_{i}\Phi\left(\frac{\Phi^{-1}(p_{i}) - \sqrts{\rho_{i}}\Phi^{-1}(1\!-\!\alpha)}{\sqrts{1-\rho_{i}}}\right) - \sum_{i = 1}^{n} \ead_{i}\lgd_{i}p_{i} \\
		& = & \sum_{i = 1}^{n} \ead_{i}\lgd_{i}\Phi\left(\frac{\Phi^{-1}(p_{i}) + \sqrts{\rho_{i}}\Phi^{-1}(\alpha)}{\sqrts{1-\rho_{i}}}\right) - \sum_{i = 1}^{n} \ead_{i}\lgd_{i}p_{i}.\IEEEyesnumber\label{eqn_cap_cred_gauss}
\end{IEEEeqnarray*} 
The last equality follows from the symmetry of the standard Gaussian density function.

Under the IRB approach developed by the Basel Committee on Banking Supervision (BCBS), regulatory capital is determined at the $99.9\%$ confidence level over a one-year horizon --- a $0.1\%$ probability that credit losses will exceed provisions and capital over the subsequent year.  In practice, it incorporates a maturity adjustment, denoted $\nu_{i}$, to account for the greater likelihood of downgrades for longer-term claims, the effects of which are stronger for claims with higher credit ratings.  Thus, regulatory capital for a portfolio comprising $n$ obligors reduces to
\begin{equation}\label{eqn_reg_cap_cred}
	\mathcal{K}_{99.9\%}(\mathcal{L}_{n}) = \sum_{i=1}^{n} \ead_{i}\lgd_{i}\nu_{i}\Phi\left(\frac{\Phi^{-1}(p_{i}) + \sqrts{\rho_{i}}\Phi^{-1}(0.999)}{\sqrts{1-\rho_{i}}}\right) - \sum_{i=1}^{n} \ead_{i}\lgd_{i}\nu_{i}p_{i}.
\end{equation}

BCBS \citeyearpar{BCBS0507} claims that the IRB approach sets regulatory capital for credit risk at a level where losses exceed it, ``on average, once in a thousand years.''  Qualifying this informal statement of probability, BCBS cautions that the $99.9\%$ confidence level was chosen because tier 2 capital ``does not have the loss absorbing capacity of tier 1'', and ``to protect against estimation error'' in model inputs as well as ``other model uncertainties.''  With provisions and capital amounting to as little as 2.0--3.0\% of EAD under the IRB approach, perhaps the claim of protection against insolvency due to credit losses at the $99.9\%$ confidence level should be interpreted as providing a margin for misspecification of the ASRF model, and not literally protection against a ``one in a thousand year'' event.  The choice of confidence level for the ASRF model may also have been influenced by the desire to produce regulatory capital requirements that are uncontroversial vis-\`a-vis Basel~I.  These qualifying remarks warn against the complacency engendered by the high confidence level chosen for the IRB approach.

In order to counter the growing problem of regulatory arbitrage under the Basel~I Accord, and encourage banks to adopt sounder risk management practices, Basel~II promotes ``better alignment of regulatory capital requirements with `economic capital' demanded by investors and counterparties'' \citep{GH06}.  With respect to credit risk, the IRB approach offers greater risk sensitivity through its use of internal (bank) estimates of unconditional PD, LGD and EAD to assess regulatory capital charges.  Despite the greater convergence between regulatory and economic capital since the implementation of Basel~II, the IRB approach is open to valid criticism:
\begin{itemize}

	\item  VaR is often criticised because it does not satisfy the subadditivity property, in general, and is therefore not a ``coherent'' risk measure \citep{ADEH99}.  However, VaR is subadditive for elliptical distributions, and the IRB approach models losses on a credit portfolio as a multivariate Gaussian distribution with default dependence described by a matrix of pairwise correlations between obligors' asset values (i.e, an elliptical distribution).  Another common criticism is that at the $\alpha$ confidence level VaR does not give any information about the severity of losses which occur with probability less than $1\!-\!\alpha$.  The margin for error and uncertainties implicit in setting a very high confidence threshold on an elliptical loss distribution arguably blunts the criticism that VaR does not account for the severity of losses beyond the confidence threshold.
	
	\item  Real-world credit portfolios do not fully satisfy Conditions~\ref{item_asymp_port} and \ref{item_single_risk_factor} above, collectively termed ``portfolio invariance'' by \citet{GMB03}.  As previously noted, credit portfolios of large, internationally active banks are typically near the asymptotic granularity of Condition~\ref{item_asymp_port}.  This assertion is supported by \citet{GL07} who apply their granularity adjustment to German bank portfolios; and \citet{RT15a} who measure the rate of convergence, in terms of number of obligors, of empirical loss distributions to the asymptotic portfolio loss distribution (i.e., the distribution of conditional expectation of portfolio percentage loss representing an infinitely fine-grained portfolio).  The Australian banks that have been granted approval to use the IRB approach certainly qualify as large, internationally active banks.  With respect to Condition~\ref{item_single_risk_factor}, many asset value models of credit risk including Moody's KMV, RiskMetrics and more advanced internal (bank) models assume that correlations are driven by several factors (e.g., geographic and industry).  But then multi-factor asset value models usually express asset values as a function of a composite factor obtained by the superposition of underlying independent risk indices \citep{BOW10}.
	
	\item  \citet{GH06} highlight that much of the literature on the procyclicality of Basel~II is written from the perspective of the formulaic minimum capital requirements, without adequate consideration of the judgemental supervisory review, which gives supervisors ``the ability to require banks to hold capital in excess of the minimum'' \citeyearpar[BCBS,][]{BCBS128}.  Furthermore, they argue that banks respond to an economic contraction (respectively, expansion) by tightening (loosening) lending standards, which serves to dampen the procyclicality of IRB capital requirements.  Nonetheless, one of the reforms introduced by Basel~III to address lessons of the financial crisis of 2007--09 promotes the build up of capital buffers through a mandatory conservation buffer and a discretionary countercyclical buffer.  The latter is intended to dampen the procyclicality of IRB capital requirements arising from internal credit ratings.  In Section~\ref{sect_cap_adeq_rpt} we look for evidence of cyclicality in the time series of unconditional PD and LGD assigned to credit exposures of the major Australian banks.
	
	\item  In a related criticism of the ``quasi'' state dependence of the model specification of the IRB approach, \citet{JRA07} argues that the concern that making regulatory capital state dependent would be procyclical and exacerbate the business cycle is overstated and perhaps unwarranted.  It is known that PD and LGD are state dependent (e.g., dependent on the state of the economy).  Recall from \eqref{eqn_cond_exp_port_loss} that the conditional PD of obligor~$i$ is given by 
\begin{equation}\label{eqn_cond_pd}
	p_{i}(y) = \Phi\left(\frac{\Phi^{-1}(p_{i}) - \sqrts{\rho_{i}}y}{\sqrts{1-\rho_{i}}}\right).
\end{equation}
Now, let $y = \Phi^{-1}(1\!-\!\alpha)$.  Then, the PD of obligor~$i$ conditional on $Y=y$ may be interpreted as the probability of default of obligor~$i$ is no greater than $p_{i}\big(\Phi^{-1}(1\!-\!\alpha)\big)$ in $(\alpha \times 100)\%$ of economic scenarios.  So, from \eqref{eqn_reg_cap_cred}, PD is state dependent in so far as $\alpha = 0.999$.  Similarly, LGD is state dependent in so far as BCBS instructs regulated institutions to provide estimates of economic-downturn LGD rates.  The rationale for this instruction is discussed in Section~\ref{sect_cap_adeq_rpt}, while the empirical analysis of Section~\ref{sect_y} recovers realisations of the single systematic risk factor describing the prevailing state of the Australian economy using downturn and through-the-cycle LGD rates.  Regulatory capital, though, serves the sole purpose of solvency assessment, so the 99.9\% confidence level and downturn LGD rates appear to be prudent choices for reducing the risk of bank failures.

	\item  BCBS has calibrated the level of correlation parameter $\rho_{i}$ in \eqref{eqn_reg_cap_cred} by asset class.  Section~\ref{sect_cap_adeq_rpt} broadly describes the asset classes defined under the IRB approach and the functions for calculating $\rho_{i}$ by IRB asset class.  The sensitivity of credit risk capital to asset correlation in the tail of the portfolio loss distribution \citep{RT15a}, draws scrutiny to the regulator specified levels of the IRB correlation parameters.  Using time series of pooled credit loss observations \citet{FITCH08} empirically estimated asset correlations for each IRB asset class.  They find that across asset classes, correlation parameters assigned under the IRB approach are conservative relative to their empirically derived estimates.  \citeauthor{FITCH08} contend that this conservatism provides a margin for error during periods of economic distress when correlations tend to spike.  Moreover, it accommodates observed geographic variations in correlation parameters.
	
	\item  In another criticism of the IRB approach, \citet{FMN01} examine model misspecification.  They simulated empirical loss distributions for a range of correlation parameters and obligor default probabilities assuming that default dependence is modelled by multivariate Gaussian and Student's $t$ distributions.  The latter admits tail dependence with fewer degrees of freedom producing stronger dependence.  Note that a multivariate $t$ distribution inherits the correlation matrix of the multivariate Gaussian distribution that it generalises.  Assuming that default dependence is modelled by a multivariate Gaussian distribution, credit losses at high quantiles of the empirical loss distribution would be seriously understated if the true model of dependence were a multivariate $t$ distribution.  In a similar exercise, \citet{RT15a} measure the sensitivity of credit risk capital to default dependence structure as modelled by the one-factor Gaussian copula and Student's $t$ copulas with Gaussian margins, and reach a similar conclusion.  \citet{FMN01} argue that asset correlations are not enough to describe default dependence, and ``[a]n assumption of multivariate normality may not model the potential extreme risk in the portfolio.''  However, they also recommend the use of historical default data to estimate true default correlations and calibrate credit risk models.  
	
	\citet{HR05} demonstrate the robustness of the Gaussian copula, which underlies the IRB approach, to model misspecification.  They simulated empirical loss distributions assuming that default dependence is modelled by the Student's $t$ copula, and then empirically estimated correlation parameters using the misspecified Gaussian copula.  While parameter estimates for asset correlation exhibit large biases, quantiles of empirical loss distributions subsequently generated by the misspecified Gaussian copula with the overestimated asset correlations no longer seriously underestimate true credit losses.  However, the economic damage wrought by the financial crisis of 2007--09 in the north Atlantic suggests that parameter estimation errors did not conveniently offset model uncertainties during the recent crisis.
		
\end{itemize}

\section{Basel~II Capital Adequacy Reporting}\label{sect_cap_adeq_rpt}
Under its implementation of Basel~II, APRA requires ADIs to assess capital adequacy for credit, market and operational risks.  ADIs determine regulatory capital for credit risk using either the standardised approach or, subject to approval, the IRB approach.  The former applies prescribed risk weights to credit exposures based on asset class and credit rating grade to arrive at an estimate of RWA.  Then, the minimum capital requirement is simply 8\% of RWA.  The standardised approach, which is an extension of Basel~I, is straightforward to administer and produces a relatively conservative estimate of regulatory capital.  The IRB approach, which implements ASRF model~\eqref{eqn_reg_cap_cred}, is a more sophisticated method requiring more input data estimated at higher precision.  Its greater complexity makes it more expensive to administer, but usually produces lower regulatory capital requirements than the standardised approach.  As a consequence, ADIs using the IRB approach may deploy their capital in pursuit of more (profitable) lending opportunities.  

Upon implementation of Basel~II in the first quarter of 2008, APRA had granted the four largest Australian banks, designated ``major'' banks, approval to use the IRB approach to capital adequacy for credit risk.  They include: Commonwealth Bank of Australia (CBA), Westpac Banking Corporation (WBC), National Australia Bank (NAB), and Australia and New Zealand Banking Group (ANZ).  WBC acquired St.~George Bank (SGB) on 1~December 2008, and CBA acquired Bank of Western Australia (BWA) on 19~December 2008.  Putting them in a global context, all four major Australian banks have been ranked in the top 20 banks in the world by market capitalisation, and top 50 by assets during 2013.  Since the implementation of Basel~II, the major banks have accounted for, on average, $74.6\%$ of total assets on the balance sheet of ADIs regulated by APRA.  Furthermore, of the regulatory capital reported by the major banks, on average,\footnote{
Average of figures for quarters ending 31~March 2008 through 30~June 2013 published by APRA \citeyearpar{AQPS13} in its quarterly issue of ADI performance statistics.
} $88.0\%$ has been assessed for credit risk, $6.9\%$ for operational risk and $4.4\%$ for market risk, with other capital charges applied by APRA accounting for the remaining $0.7\%$.\footnote{
Data are reported for the major Australian banks, in aggregate, so as not to violate confidentiality agreements.
}

ADIs lodge their statutory returns with APRA using a secure electronic data submission system \citeyearpar[APRA,][]{APRA1406}.  A return is a collection of related forms covering the same reporting period.  A form is a dataset containing information on a specific topic (e.g., capital, risk class/ sub-class, financial statement, etc.).  For reference purposes, forms in spreadsheet format and instructions are available at \url{www.apra.gov.au}.  Given the volume of data processed in preparing statutory returns, the major Australian banks automate their electronic data submissions using XBRL (eXtensible Business Reporting Language) --- an XML-based language for preparing, publishing, extracting and exchanging business and financial information.  Once forms are validated and returns submitted, data are stored in APRA's data warehouse.  We use SQL programming to fetch and aggregate data reported to APRA by the major banks on a quarterly basis from 31~March 2008 through 30~June 2013.\footnote{
NAB did not adopt the IRB approach to capital adequacy for credit risk until the second quarter of 2008.  Therefore, in measuring the effect of the financial crisis, we omit NAB from our major banks' aggregate for the quarter ending 31~March 2008.
}  Specifically, data reported on the statement of financial performance, capital adequacy form, and IRB credit risk forms.\footnote{
For a number of quarters after the acquisitions of BWA and SGB by CBA and WBC, respectively, BWA and SGB reported credit risk using the IRB approach in parallel with the consolidated reporting of CBA and WBC, which determined RWA for credit risk in the banking books of BWA and SGB using the standardised approach.  In our analysis we have included the parallel IRB credit risk forms submitted by BWA and SGB, and deducted the corresponding RWA, as determined by the standardised approach, from the consolidated RWA reported by CBA and WBC, respectively.
}

\begin{figure}[!t]
	\centering
	\caption[EAD and RWA of IRB credit exposures of the major Australian banks by sector]{EAD and RWA, respectively, of IRB credit exposures of the major Australian banks are decomposed by sector (business, government and household).\protect\\
		{\footnotesize Source: Australian Prudential Regulation Authority.}}
	\label{fig_irb_ead_rwa}
	\scalebox{0.51}{
	\begin{tabular}{c c}
		\input{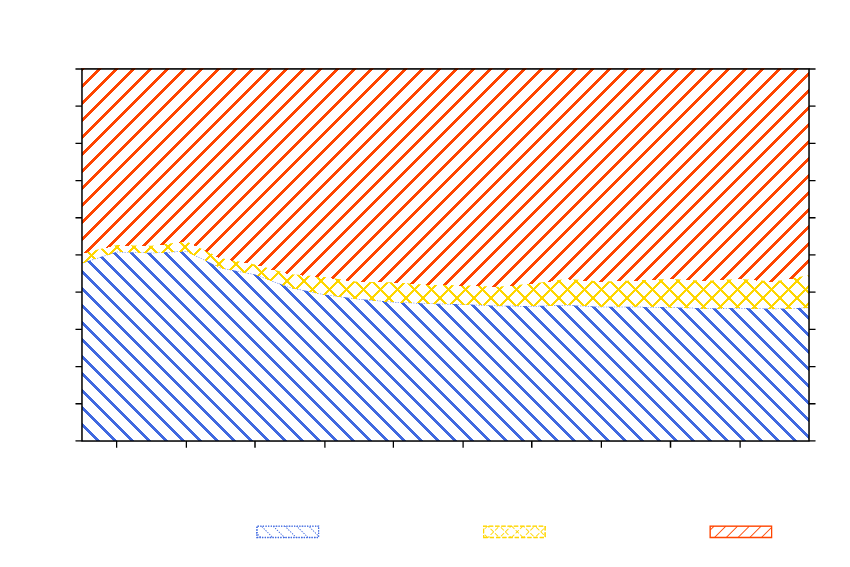}	& \input{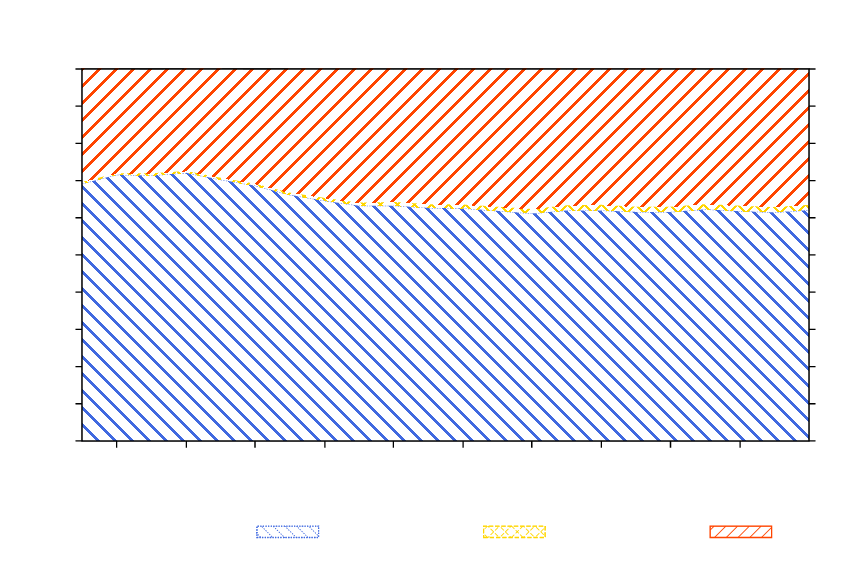}
	\end{tabular}
	}
\end{figure}

Banking book exposures are reported on credit risk forms\footnote{
Reporting forms and instructions for ADIs available at \url{www.apra.gov.au}: ARF\_113\_1A, ARF\_113\_1B, ARF\_113\_1C, ARF\_113\_1D, ARF\_113\_3A, ARF\_113\_3B, ARF\_113\_3C, and ARF\_113\_3D.
} by IRB asset class: corporate (non-financial), small- and medium-sized enterprises (SME), bank, sovereign, residential mortgages, retail qualified revolving, and other retail\footnote{
APRA \citeyearpar{APS113} provides a full definition of IRB asset classes.
}.  For presentation purposes we merge IRB asset classes, and report credit exposures as business, government or household.  Henceforth, we refer to these banking book exposures as IRB credit exposures.  Figure~\ref{fig_irb_ead_rwa} decomposes EAD and RWA, respectively, of IRB credit exposures into business, government and household sectors.  Since the implementation of Basel~II, RWA for credit risk reported by the major Australian banks has been divided, on average, 71.9/28.1 between IRB credit exposures and other banking book exposures.  The market dominance of the major banks, coupled with the concentration of their regulatory capital held against unexpected losses on IRB credit exposures, convey the significance of the ASRF model in protecting the Australian banking sector against insolvency.  Accordingly, we contend that, while focussing exclusively on IRB credit exposures of the major banks, our empirical analysis draws a representative sample of the credit risk assumed by the Australian banking sector.

\afterpage{
\begin{landscape}
\begin{figure}[!t]
	\caption[Risk characteristics of credit portfolios of the major Australian banks by sector]{Risk characteristics of credit portfolios (IRB credit exposures) of the major Australian banks, rebalanced quarterly, by sector (business, government and household), as well as for the whole portfolio.  Credit risk variables and parameters include: exposure-weighted unconditional PD, economic-downturn LGD, asset correlation and maturity adjustment.\protect\\
		{\footnotesize Sources: Australian Prudential Regulation Authority.}}
	\label{fig_port_char_sector}
	\centering
	\scalebox{0.63}{
	\begin{tabular}{cc}
		\input{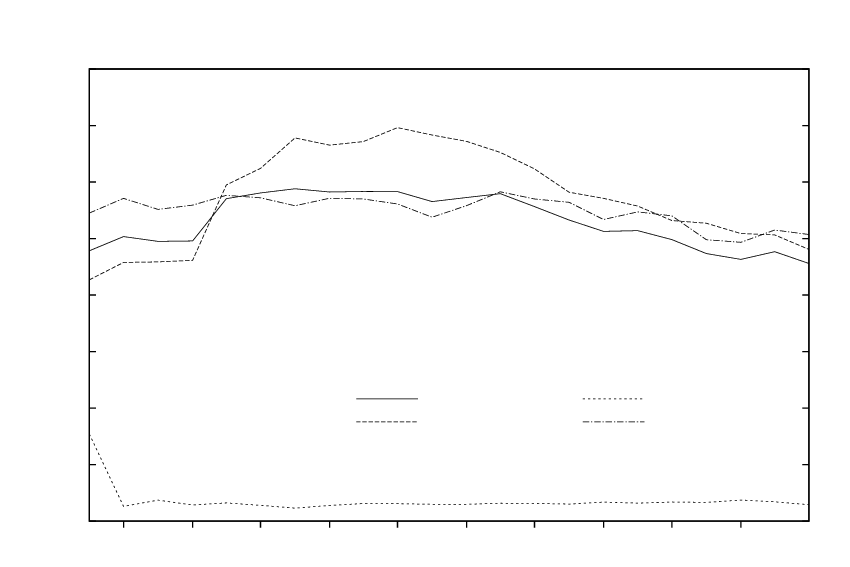}	& \input{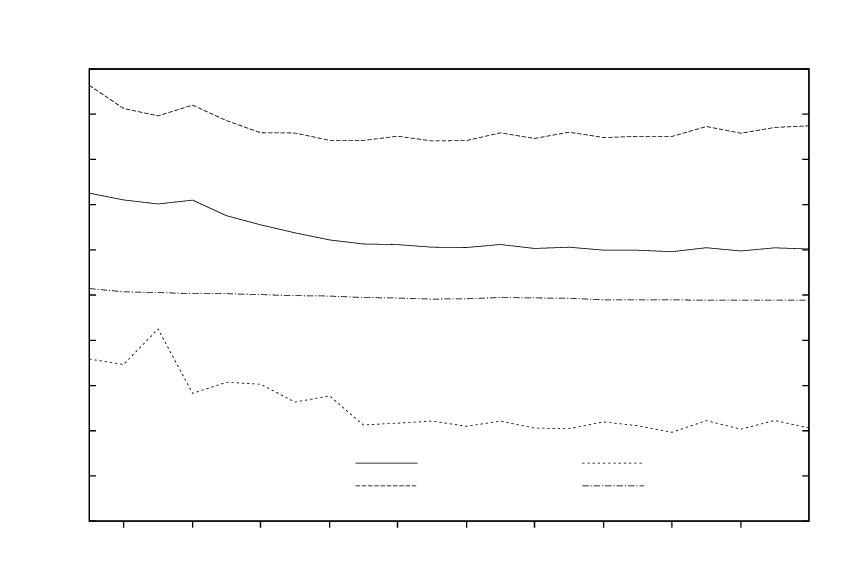}	\\
							&					\\
		\input{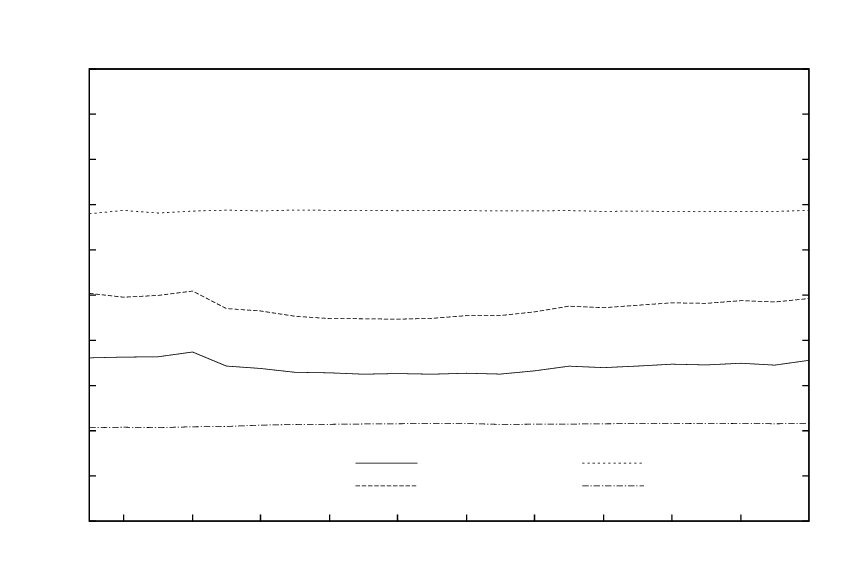}	& \input{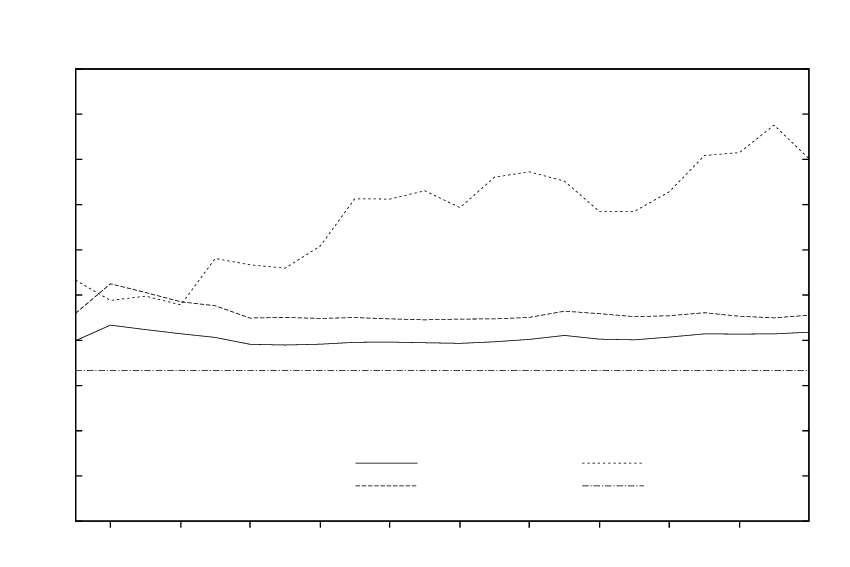}
	\end{tabular}
	}
\end{figure}
\end{landscape}
}

Under the IRB approach, ADIs assign their on- and off-balance sheet credit exposures to internally-defined obligor grades reflecting PD bands, and LGD bands.  Between 2008 and 2013 RWA of IRB credit exposures held in the banking book of the major Australian banks has been divided, on average, 75.4/24.6 between on-balance sheet assets and off-balance sheet exposures.  EAD, RWA, expected loss, and exposure weighted LGD, unconditional PD, maturity and firm size are reported for each obligor grade.  We assign IRB credit exposures reported by the major banks to standardised PD bands (i.e., consistent across the major banks), and calculate risk parameters characterising each of these standardised obligor grades.  The ASRF model incorporates a maturity adjustment, which is a function of maturity and unconditional PD, for business and government credit exposures.  Asset correlation is constant for residential mortgages and retail qualified revolving credit exposures; a function of unconditional PD for corporate, bank, sovereign and other retail credit exposures; and a function of firm size and unconditional PD for SME credit exposures \citeyearpar[BCBS,][]{BCBS0507}.  We construct a commingled credit portfolio by pooling obligor grades across IRB asset classes.  Figure~\ref{fig_port_char_sector} plots exposure-weighted PD, LGD, asset correlation and maturity adjustment for commingled credit portfolios, rebalance quarterly, by sector (business, government and household), as well as for the whole portfolio.

Prudential standards published by APRA instruct ADIs to assign one-year unconditional (through-the-cycle) PDs and economic-downturn (stressed) LGD rates to IRB credit exposures.  If the major banks were reporting unconditional PDs, one would expect that temporal variations in reported PDs would be explained exclusively by changes in portfolio composition.  In reality though, the prevailing state of the economy likely affects reported PDs to a degree.  That is, reported PDs are the result of perturbations of through-the-cycle estimates by point-in-time influences.  Figure~\ref{fig_port_char_sector} shows the variation in reported exposure-weighted PD for the whole portfolio ranging from 0.91\% to 1.18\%.  Assuming that portfolio composition was unchanged between 2008 and 2013, and using an exposure-weighted asset correlation of 16.8\%,\footnote{
The exposure-weighted asset correlation of the IRB credit exposures of the major Australian banks averaged 0.168 for quarters ending 31~March 2008 through 30~June 2013.
} the corresponding range of realisations of the single systematic risk factor describing the state of the economy is less than one-quarter of one standard deviation.  This is a fraction of the range of realisations of the single systematic risk factor recovered from the ASRF model in Section~\ref{sect_y}.  Accordingly, we don't believe that our conclusions are materially effected by point-in-time influences on reported PDs.

During an economic downturn losses on defaults are likely to be higher than under ``normal'' economic conditions.  In developing the IRB approach to capital adequacy for credit risk, BCBS considered implementing a function that translates average, or through-the-cycle, LGD rates into downturn LGD rates, similar to the function derived by \citet{VO02}, which transforms unconditional PDs into conditional PDs.  In principle, this function could depend on several factors including the state of the economy, magnitude of the average LGD, and asset class of the exposure.  Given the evolving nature of LGD modelling and practices for estimating LGD at the time, BCBS elected to instruct ADIs to provide estimates of downturn LGD rates \citeyearpar[BCBS,][]{BCBS0507}.  There is no inconsistency here when assessing regulatory capital charges, since capital is held against unexpected losses arising under recessionary conditions.  However, using downturn LGD rates to recover realisations of the single systematic risk factor from the ASRF model yields a prevailing state of the economy that is more expansionary than would be recovered using point-in-time LGD rates, assuming that the economy was not experiencing a (severe) downturn.  The empirical analysis of Section~\ref{sect_y} addresses this complication by recovering realisations of the single systematic risk factor describing the prevailing state of the Australian economy using downturn and through-the-cycle LGD rates.  Note that distance-to-default, which measures the level of capitalisation, reflects the capacity to absorb credit losses under stressed economic conditions.  So, it is appropriate to use downturn LGD rates when translating realisations of the single systematic risk factor recovered from the ASRF model into estimates of distance-to-default.

\begin{figure}[!t]
	\centering
	\caption[Risk-based capital ratio of the major Australian banks]{Risk-based capital ratio (capital / RWA) of the major Australian banks decomposed into tier 1 and tier 2 capital.\protect\\
		{\footnotesize Source: Australian Prudential Regulation Authority.}} 
	\label{fig_cap_ratio}
	\scalebox{0.80}{
		\input{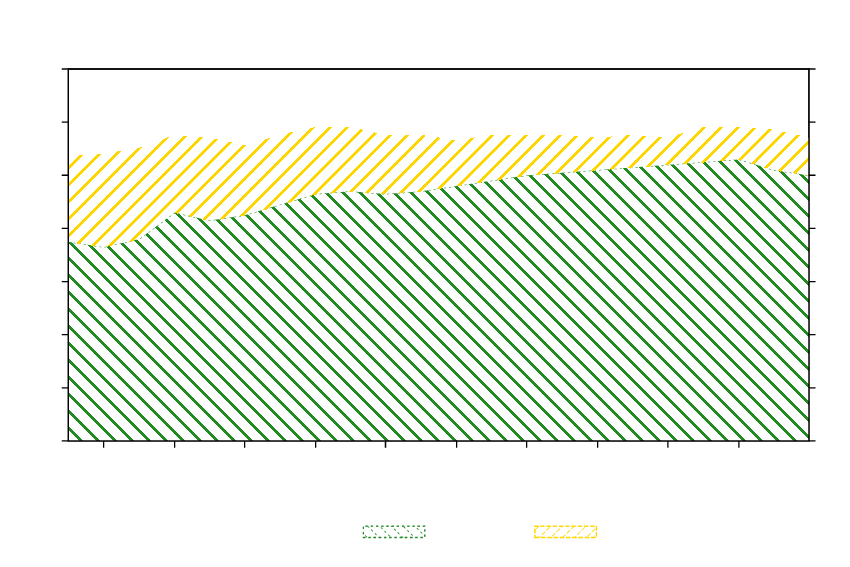}
	}
\end{figure}

RWA, capital and provisions are reported on the capital adequacy form.\footnote{
Reporting form ARF\_110\_0\_2 and instructions for ADIs available at \url{www.apra.gov.au}.
}  RWA are reported by risk class, and within the credit risk class, for IRB asset classes and the standardised approach.  These data are aggregated across capital adequacy forms submitted by the major Australian banks.  Note that the minimum capital requirement is simply 8\% of RWA.  Then, subject to a minimum 8\% of RWA, APRA sets a prudential capital ratio for each ADI, and an ADI typically holds a capital buffer above its prudential capital requirement \citeyearpar[APRA,][]{APRA0712}.  Figure~\ref{fig_cap_ratio} decomposes the aggregate capital base of the major banks, measured as a percentage of RWA, into tier~1 and tier~2 capital.

Finally, we take credit losses as charges for bad and doubtful debts reported on the statement of financial performance,\footnote{
Reporting form ARF\_330\_0\_C and instructions for ADIs available at \url{www.apra.gov.au}.
} or income statement.  Credit losses are aggregated across income statements reported to APRA by the major banks.

\section{Recent Performance of the Australian Economy}\label{sect_perf_au_econ}
In conducting our evaluation of the model specification of the Basel~II IRB approach, we use internal bank data collected by APRA to solve for realisations of the single systematic risk factor describing the prevailing state of the Australian economy.  These readings of the Australian banking sector are then compared with signals from macroeconomic indicators and financial statistics\footnote{
Measures of financial position (stocks) and performance (flows), national and company level.
} to render our assessment.  Here, we describe the performance of the Australian economy over the past decade, a period which includes the financial crisis of 2007--09, also known as the global financial crisis.  While the crisis precipitated the worst global recession since the Great Depression, its effects were not evenly felt across the globe.  In order to contextualise the Australian economy we compare its performance with that of the United States and United Kingdom, economies that experienced the full force of the recent crisis.  The resilience of the Australian economy to the crisis is evident from a review of macroeconomic indicators and financial statistics that portend or reflect credit stresses: real GDP growth, unemployment rate, house prices, and return on equity of the banking sector.

A number of explanations for the resilience of the Australian economy have been proffered:
\afterpage{
\begin{landscape}
\begin{figure}[!p]
	\caption[Resilience of the Australian economy to the financial crisis of 2007--09 vis-\`a-vis the economies of the US and UK]{Resilience of the Australian economy to the financial crisis of 2007--09.  Its performance is compared with that of the United States and United Kingdom on macroeconomic indicators and financial statistics: real GDP growth, unemployment rate, house price index, and return on equity of the banking sector.\protect\\
		{\footnotesize Sources: Australian Bureau of Statistics; Australian Prudential Regulation Authority; US Bureau of Economic Analysis; US Bureau of Labor Statistics; US Federal Reserve Economic Data; Standard \& Poor's;  UK Office of National Statistics.}} 
	\label{fig_macro_econ}
	\centering
	\scalebox{0.67}{
	\begin{tabular}{cc}
		\input{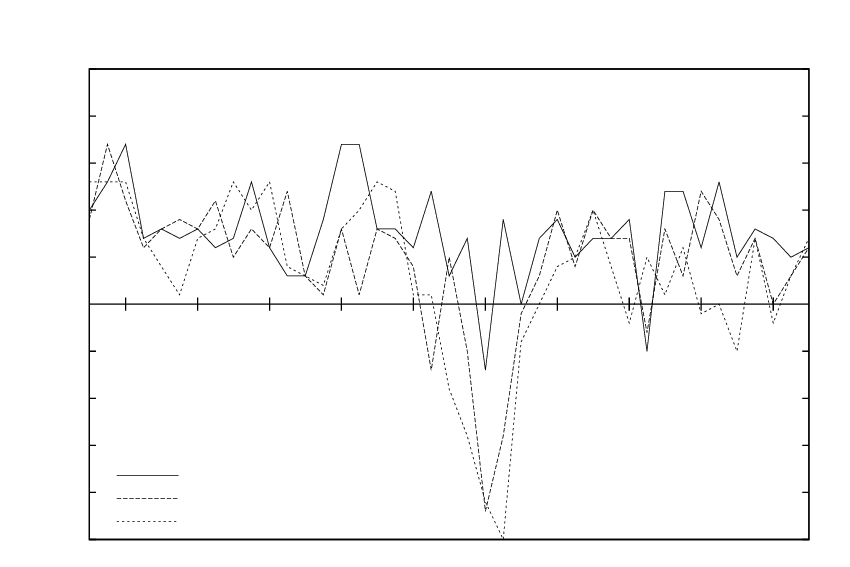}		& \input{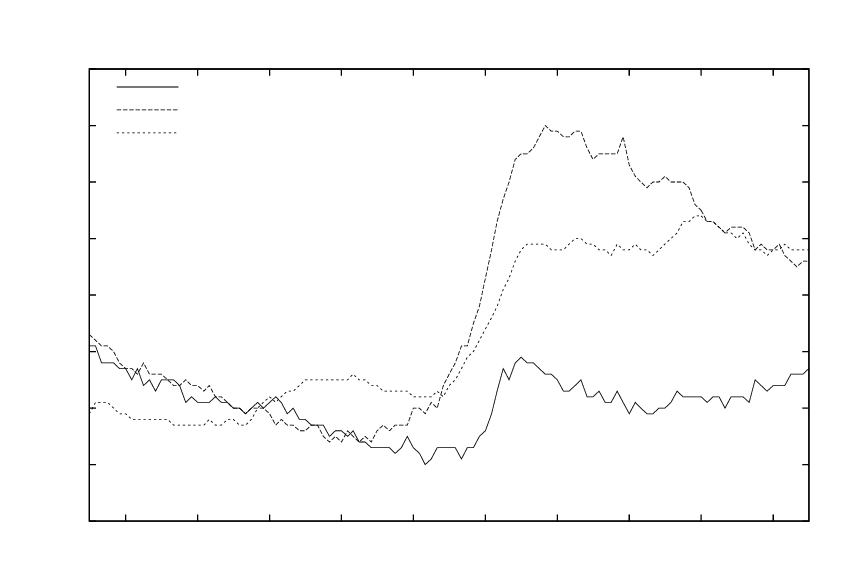}	\\
		\input{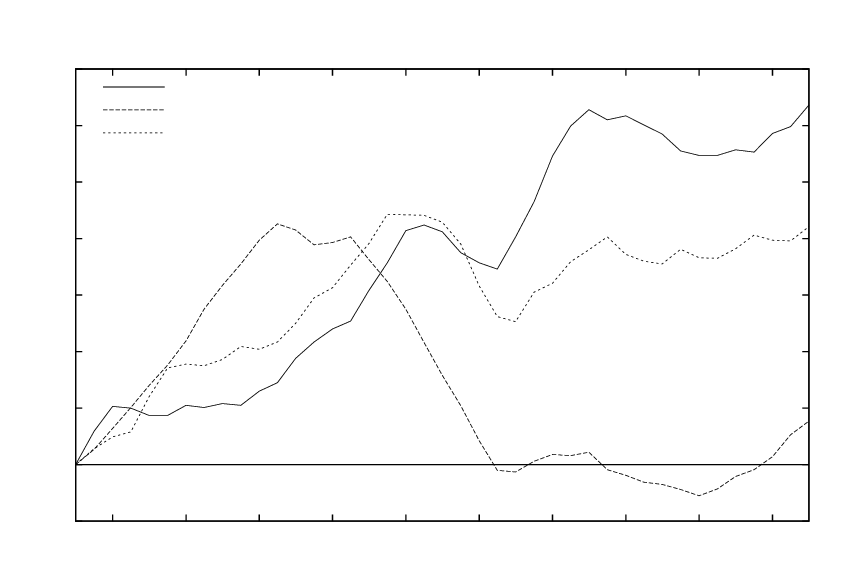}		& \input{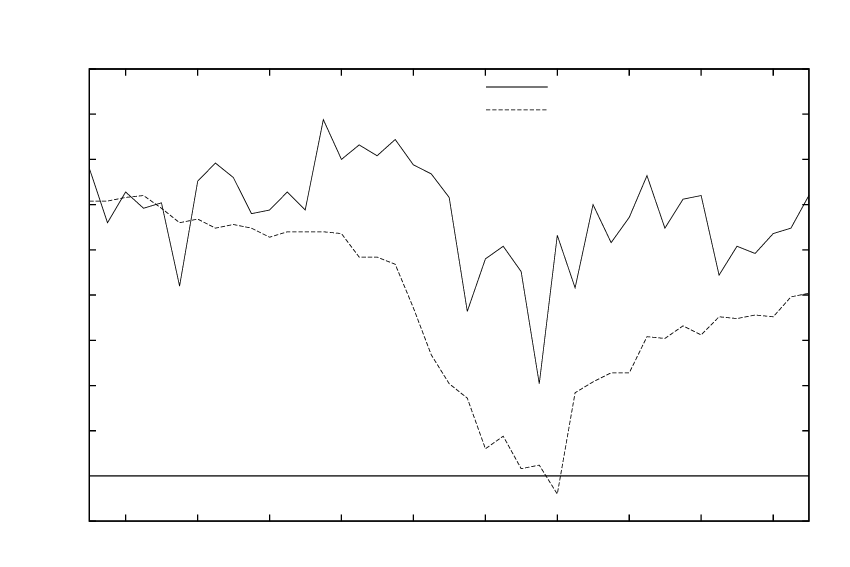}
	\end{tabular}
	}
\end{figure}
\end{landscape}
}
\begin{itemize}
	\item  Since the mid 2000s Australia has benefited from a favourable movement in terms of trade driven by the strong global demand for commodities, much of it coming from Asia.  An appreciating foreign exchange rate over the same period has been a key factor in the relatively smooth adjustment of the economy to the increase in terms of trade \citep{BKPR13}.  Inflation has been consistent with the target set by the Reserve Bank of Australia (RBA), unemployment has remained relatively low, and economic growth has mostly been around trend --- the economy registered a mild contraction in the fourth quarter of 2008, but did not experience a recession.  These economic fundamentals have shielded Australia from the financial crisis.
	\item  Residential mortgages, typically floating rate with full recourse to the borrower, have accounted for 50--60\% of loans on the balance sheet of Australian banks over the past decade.  With the vast bulk of residential mortgages originated by the banks and held to maturity, the risks associated with the banks' residential mortgage portfolios are comparatively small.  Other notable structural factors of the Australian financial sector which serve to constrain excessive risk taking include: the government's ``four pillars'' policy precluding mergers between the four major banks, which dominate the Australian financial sector; and the ability of the Australian banks to pay fully-franked dividends to shareholders \citep{LD13}.  The attention to risk quantification and management by the major banks, in order to achieve Basel~II IRB approval in 2008, arguably helped discourage excessive risk-taking too \citep{BD10}.
	
	\item  APRA, which is responsible for the regulation of deposit-taking institutions, insurance companies and superannuation funds, distinguishes prudential supervision in Australia by its active, risk-based approach \citep{LD13}.  BCBS \citeyearpar{BCBS230} calls for greater focus on risk-based supervision in its core principles for effective banking supervision:  ``[t]his risk-based process targets supervisory resources where they can be utilised to the best effect, focussing on outcomes as well as processes, moving beyond passive assessment of compliance rules.''  There were significant failures of non-bank financial companies during the financial crisis with investors, rather than taxpayers, bearing the losses.  The list of failed companies includes: Absolute Capital, Allco Finance Group, Babcock and Brown, Opes Prime, RAMS Home Loans and Storm Financial.  Australia's banking sector, on the other hand, experienced no failures.  Indeed, while banks' profitability declined during the crisis, it remained quite healthy.  Also, the major Australian banks maintained their AA credit rating through the crisis \citep{BD10, DAT10}.
	
	\item  In October 2008 the Australian Government introduced two schemes, announced with a three-year duration, to guarantee bank funding.  Under the Financial Claims Scheme, all deposits up to \$1 million with locally-incorporated ADIs were automatically guaranteed by the government with no fee payable.  Under the Funding Guarantee Scheme, the government provided a guarantee, for a fee, on deposits over \$1 million and wholesale funding with maturity out to five years.  The RBA believes that these schemes achieved their goal of maintaining public confidence in the Australian banking sector \citep{SSCE11}.
	
	\item  At the onset of the financial crisis, in October 2007, the RBA sought to restore liquidity to dysfunctional credit markets by expanding the range of securities it would accept as collateral for repurchase agreements to include residential mortgage-backed securities and asset-backed commercial paper \citep{DG07}.  Then, as the crisis spread to the real economy, the RBA slashed its target cash rate from $7.25\%$ in August 2008 to $3.0\%$ in April 2009.  Importantly, much of the monetary policy easing was passed through to borrowers.  With most household and business loans in Australia being variable, lower interest rates translated into higher disposable incomes.  Falling interest rates coupled with rising incomes improved housing affordability somewhat, and helped avert a sharp correction in the housing market, which had boomed over the previous decade \citep{MM11}.
	
	\item  Between October 2008 and February 2009 the Australian Government announced substantial fiscal stimulus packages: \$10.4~billion Economic Security Strategy; \$15.2~billion Council of Australian Governments funding package; \$4.7~billion Nation Building package; and \$42~billion Nation Building and Jobs Plan.  The Treasury estimates that, absent the fiscal stimulus, GDP growth would have been negative for three consecutive quarters \citep{MM11}.
	
	\item  Finally, the ``lucky'' country was probably not without a dose of good fortune.
\end{itemize}

The charts in Figure~\ref{fig_macro_econ} clearly indicate that the Australian economy was largely cushioned from the adverse effects of the recent crisis.  Putting the mild contraction experienced by the Australian economy during the financial crisis of 2007--09 in an historical context, we compare it with Australia's most recent recession in 1990.  During the financial crisis real GDP growth in Australia fell 0.7\% in the fourth quarter of 2008, and unemployment peaked at 5.9\% in June 2009.  By comparison, during the 1990 recession the Australian economy contracted 1.7\% and unemployment rose to 10.8\%.  A number of financial institutions failed including the State Bank of Victoria, the State Bank of South Australia, the largest credit union (Teachers Credit Union of Western Australia), the second largest building society (Pyramid Building Society), and several merchant banks.  Moreover, two of the four major Australian banks incurred heavy losses and had to be recapitalised.  The 1990 recession is considered similar in magnitude to those of 1974 and 1961, but not as severe as the recession of 1982 when the economy contracted 3.7\% and unemployment rose to 10.5\% \citep{MI06}.

\section{Measurements from the ASRF Model of the Australian Banking Sector}\label{sect_asrf_au_banks}
The ASRF model prescribed under the Basel~II IRB approach is an asset value factor model.  As noted in Section~\ref{sect_model_spec_irb}, other asset value models include Moody's KMV, RiskMetrics, and most internal bank models.  They too are generally factor models.  We believe that our empirical findings on the ASRF model are applicable to the broader class of asset value factor models of credit risk.  

Applying the ASRF model, with appropriate substitutions, to internal bank data collected by APRA, our empirical analysis generates time series of: (\textit{i})~the single systematic risk factor describing the prevailing state of Australia's economy; and (\textit{ii})~distance-to-default measuring the capacity of its banking sector to absorb credit losses.  Then, comparing these time series with macroeconomic indicators, financial statistics and external credit ratings we render a fundamental assessment of the model specification of the IRB approach.  Furthermore, since the depths of the financial crisis of 2007--09 were reached after the implementation of Basel~II, this time series analysis measures the impact of the crisis on the Australian banking sector.  Finally, in a variation on the evaluation of distance-to-default, reverse stress testing explores stress events that would trigger material supervisory intervention.  Again, access to internal bank data collected by the prudential regulator distinguishes our research from other empirical studies on the IRB approach and recent crisis.

Taking measurements from the ASRF model of the Australian banking sector requires internal bank data collected by APRA under Basel~II.  Data reported on IRB credit risk forms by the major banks are available since the first quarter of 2008 on a quarterly basis.  Recall that IRB credit risk forms assign credit exposures to an obligor grade identifying a PD band within an IRB asset class.  Pooling obligor grades, reported as at the end of a given quarter, across IRB asset classes constitutes a commingled credit portfolio in which each obligor grade is represented as a single credit.  The quarterly observations between 31~March 2008 and 30~June 2013 establish a time series of commingled credit portfolios. 

Recall that $\ead_{i}(t)$, $\lgd_{i}(t)$, $\nu_{i}(t)$, $p_{i}(t)$ and $\rho_{i}(t)$ denote the EAD, LGD, maturity adjustment, unconditional PD and asset correlation assigned to credit~$i$ as at the end of quarter $t$.  They are input to the ASRF model, which makes projections of losses on IRB credit exposures over the subsequent four-quarter interval $[t\!+\!1,t\!+\!4]$.  As noted in Section~\ref{sect_cap_adeq_rpt}, these parameters and variables are reported on IRB credit risk forms submitted to APRA by the majors banks on a quarterly basis.

\subsection{Prevailing State of the Australian Economy}\label{sect_y}
The model specification of the IRB approach assesses regulatory capital charges by calculating the expectation of credit losses conditional on a realisation of the single systematic risk factor.  Substituting losses incurred on IRB credit exposures into the ASRF model, we recover realisations of the single systematic risk factor describing states of the Australian economy experienced since the implementation of Basel~II.  Then, applying the standard Gaussian distribution function yields the quantile of the distribution of economic scenarios.

For any one-year horizon over which the ASRF model makes projections of credit losses, we choose to associate the model inputs with the realisation of the single systematic risk factor as at the midpoint of the corresponding risk measurement horizon.  So, measuring time in quarterly increments, realisation~$y(t)$ of systematic risk factor~$Y$ describing the state of the economy as at the end of quarter~$t$ is associated with credit losses projected over the time interval $[t\!-\!1,t\!+\!2]$ using credit risk parameters and variables reported as at the end of quarter~$t\!-\!2$.

Suppose that for each quarter in the time series, there is a total of $r$ banking book exposures subject to credit risk, of which $n$ IRB credit exposures are held in the aforementioned commingled portfolio, where $n \leq r$.  Denote by $\mathcal{R}_{n}(t)$, $\mathcal{R}_{r}(t)$ and $\mathcal{R}(t)$ the RWA of IRB credit exposures, RWA for credit risk and total RWA, respectively, as at the end of quarter~$t$.  Let $\mathcal{L}_{r}(s)$ be the total credit losses incurred during quarter~$s$, which is reported on the statement of financial performance.  We do not observe $\mathcal{L}_{n}(s)$, losses incurred on IRB credit exposures during quarter~$s$.  Therefore, we choose to allocate total credit losses incurred during the four-quarter interval $[s\!-\!1, {s\!+\!2}]$ between IRB credit exposures and other banking book exposures in proportion to RWA as at the end of the quarter~$t\!-\!2$.  So, losses on IRB credit exposures incurred during quarters $s\!-\!1$, $s$, $s\!+\!1$ and $s\!+\!2$ is given by 
\begin{equation}
	\frac{\mathcal{R}_{n}(t\!-\!2)}{\mathcal{R}_{r}(t\!-\!2)}\big( \mathcal{L}_{r}(s\!-\!1) + \mathcal{L}_{r}(s) + \mathcal{L}_{r}(s\!+\!1) + \mathcal{L}_{r}(s\!+\!2) \big).
\end{equation}

In taking measurements from the ASRF model of the Australian banking sector, we assume that there is no delay in the recognition of bad debts.  (Later, we allow for delays in the recognition of bad debts.)  Setting $s = t$, projected credit losses over any one-year horizon are compared with credit losses incurred during the same one-year interval.  Then, applying the formula for conditional expectation of portfolio credit losses,
\begin{IEEEeqnarray*}{rCl}
	\IEEEeqnarraymulticol{3}{l}{\frac{\mathcal{R}_{n}(t\!-\!2)}{\mathcal{R}_{r}(t\!-\!2)}\big( \mathcal{L}_{r}(t\!-\!1) + \mathcal{L}_{r}(t) + \mathcal{L}_{r}(t\!+\!1) + \mathcal{L}_{r}(t\!+\!2) \big)} \\
	\qquad & = & \sum_{i=1}^{n} \ead_{i}(t\!-\!2)\lgd_{i}(t\!-\!2)\nu_{i}(t\!-\!2) \Phi\left(\frac{\Phi^{-1}\big(p_{i}(t\!-\!2)\big) - \sqrts{\rho_{i}(t\!-\!2)}y(t)}{\sqrts{1-\rho_{i}(t\!-\!2)}}\right),\label{eqn_cred_loss_cond_exp}\IEEEyesnumber
\end{IEEEeqnarray*}
we solve for realisation~$y(t)$ of systematic risk factor~$Y$.\footnote{
We use the GNU Scientific Library routine \texttt{gsl\_root\_fsolver\_falsepos}, which implements the \textit{false position algorithm}, to solve for $y(t)$.  A root bracketing algorithm based on linear interpolation, its convergence is linear, but it is usually faster than the \textit{bisection method}.
}  Repeating this static analysis for each quarter~$t$ from 30~September 2008 through 31~December 2012 generates a time series of systematic risk factor~$Y$.  The ASRF model assumes that the conditional expectation of portfolio credit losses rises as the economy deteriorates --- a strictly decreasing function of the single systematic risk factor.  By convention the $\alpha$ quantile of the distribution of ${\E[L_{n} \given Y]}$ is associated with the $1\!-\!\alpha$ quantile of the distribution of~$Y$.  Hence, confidence level~$\alpha(t)$ corresponding to realisation~$y(t)$ of systematic risk factor~$Y$ is given by
\begin{equation}
	\alpha(t) = 1 - \Phi\big(y(t)\big).
\end{equation}

As discussed in Section~\ref{sect_cap_adeq_rpt}, prudential standards published by APRA instruct ADIs to assign economic downturn (stressed) LGD rates to IRB credit exposures.  Denote by $\underline{\lgd_{i}}(t)$ the downturn LGD assigned to credit~$i$ as at the end of quarter~$t$.  Setting $\lgd_{i}(t) = \underline{\lgd_{i}}(t)$ for $i = 1, \ldots, n$ in \eqref{eqn_cred_loss_cond_exp} and solving for realisation~$y(t)$ of systematic risk factor~$Y$ yields a prevailing state of the economy that is more expansionary than would be recovered using point-in-time LGD rates, assuming that the economy was not experiencing a (severe) downturn.  In solving for realisations of the single systematic risk factor describing the prevailing state of the economy, we also examine the effect of adjustments to downturn LGD rates.  Recognising that banks may not be able to produce internal estimates of downturn LGD, the \citet{OCC0711} proposed a ``linear supervisory mapping function'' for translating through-the-cycle LGD rates into downturn LGD rates: ${\underline{\lgd_{i}}(t) = 0.08 + 0.92\bar{\lgd_{i}}(t)}$, where $\bar{\lgd_{i}}(t)$ is the through-the-cycle LGD assigned to credit~$i$ at the end of quarter~$t$.  The adjustment varies between zero and eight percent, decreasing linearly from low to high through-the-cycle LGD rates.  Using historical LGD rates from their database, Moody's derived an adjustment method to meet the downturn LGD requirement imposed under the IRB approach.  Their adjustment to through-the-cycle LGD rates depends on the seniority of the obligations, with the magnitude also varying from zero to eight percent \citep{EK07}.  We demonstrate the effect of adjustments to downturn LGD rates by solving \eqref{eqn_cred_loss_cond_exp} using four definitions of LGD: $\lgd_{i}(t) = \underline{\lgd_{i}}(t)$, $\lgd_{i}(t) = \bar{\lgd_{i}}(t)$, $\lgd_{i}(t) = \tfrac{1}{3}\underline{\lgd_{i}}(t)$, and $\lgd_{i}(t) = \tfrac{1}{2}\underline{\lgd_{i}}(t)$ for $i = 1, \ldots, n$.

\begin{figure}[!t]
	\centering
	\caption[Realisations of the single systematic risk factor describing the prevailing state of the Australian economy, and the effect of adjustments to downturn LGD]{Realisations of the single systematic risk factor describing the prevailing state of the Australian economy, and the effect of adjustments to downturn LGD.  Confidence level~$\alpha$ is the probability of the state of the economy being better than the economic scenario described by realisation~$y$.  For credits $i = 1, \ldots, n$ constituting the commingled portfolio at the end of quarter~$t$:\protect\\
	(\textit{a}) $\lgd_{i}(t) = \underline{\lgd_{i}}(t)$, where $\underline{\lgd_{i}}(t)$ is downturn LGD.\protect\\
	(\textit{b}) $\lgd_{i}(t) = \bar{\lgd_{i}}(t)$, where $\bar{\lgd_{i}}(t)$ is through-the-cycle LGD and \protect\\\hspace*{77pt}${\underline{\lgd_{i}}(t) = 0.08 + 0.92\bar{\lgd_{i}}(t)}$.\protect\\
	(\textit{c}) $\lgd_{i}(t) = \tfrac{1}{3}\underline{\lgd_{i}}(t)$.\protect\\
	(\textit{d}) $\lgd_{i}(t) = \tfrac{1}{2}\underline{\lgd_{i}}(t)$.}
	\label{fig_y_alpha_lgd}
	\scalebox{0.90}{
	\begin{tabular}{c}
		\input{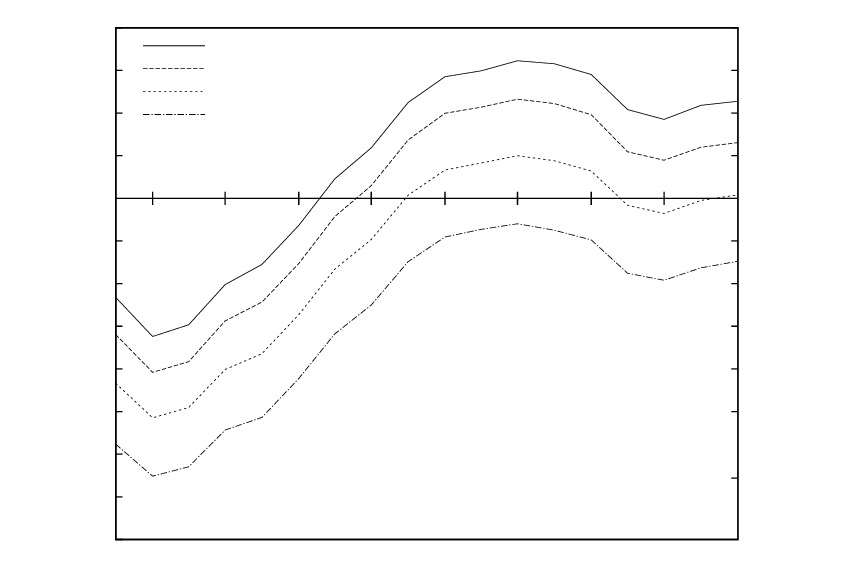}
	\end{tabular}
	}
\end{figure}

Figure~\ref{fig_y_alpha_lgd} plots realisations of the single systematic risk factor describing the prevailing state of the economy.  On the basis of downturn LGD rates, as reported by the major banks, the economic shock imparted by the financial crisis of 2007--09 propagated through the Australian banking system inflicting credit losses incurred, on average, once every five years.  That is, at the depths of the crisis, the prevailing state of the Australian economy as recovered from \eqref{eqn_cred_loss_cond_exp} is summarised by realisation $y = -0.81$ on 31~December 2008.  Appealing to Gordy's proposition, $y = -0.81$ translates into the $79.1\%$ quantile of the portfolio loss distribution --- credit losses incurred were no greater than expected in $79.1\%$ of economic scenarios, which we characterise informally as (approximately) a one-in-five-year event.  We argue that the single systematic risk factor is a germane measure of the impact of the financial crisis, because the IRB approach chooses realisation~$y = -3.090$ of systematic risk factor~$Y$ to assess regulatory capital charges that are expected to absorb credit losses in 99.9\% of economic scenarios.  Bear in mind, though, that measurements from the ASRF model describing the prevailing state of the Australian economy capture policy responses of government departments and agencies (i.e., The Treasury, RBA and APRA) designed to mitigate the recent crisis.  Naturally, it is not possible to isolate these policy responses in order to ascribe a value to their mitigating effects.

Translating downturn LGD rates into through-the-cycle LGD rates using the linear supervisory mapping function proposed by \citet{OCC0711}, realisations of the single systematic risk factor recovered from \eqref{eqn_cred_loss_cond_exp} indicate that the recent crisis inflicted credit losses on the Australian banking system incurred, on average, once every seven years.  Given that the Australian economy contracted during the financial crisis, and assuming that through-the-cycle LGD rates obtained from downturn LGD rates using the linear supervisory mapping function are good estimates, we would argue that the shock experienced by the Australian economy at the depths of the crisis was between a one-in-five-year and one-in-seven-year event.  This seems consistent with the perception that Australian banks weathered the recent crisis quite well, reporting good profitability and high capital ratios, and maintaining strong credit ratings.

\begin{figure}[!t]
	\centering
	\caption[Realisations of the single systematic risk factor and real GDP growth of the Australian economy]{Realisations $y(t)$ of systematic risk factor $Y$ describing the prevailing state of the Australian economy and real GDP growth, y-o-y.  The time series exhibit a moderately strong correlation (+0.60).\protect\\
		{\footnotesize Source: Australian Bureau of Statistics.}}
	\label{fig_y_gdp}
	\scalebox{0.90}{
	\begin{tabular}{c}
		\input{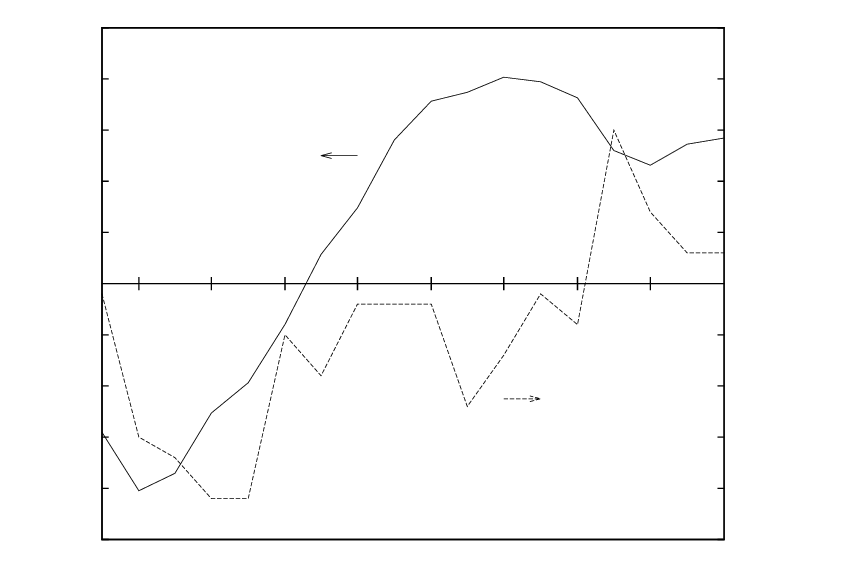}
	\end{tabular}
	}
\end{figure}

For reference, Figure~\ref{fig_y_alpha_lgd} also plots realisations of the single systematic risk factor recovered from \eqref{eqn_cred_loss_cond_exp} using LGD estimates that are one-third less than, and half of downturn LGD rates.  Under these assumptions for point-in-time LGD at the depths of the crisis, the shock experience by the Australian economy was a one-in-ten-year and one-in-twenty-year event, respectively.  In the analysis that follows we use the time series of the single systematic risk factor recovered from \eqref{eqn_cred_loss_cond_exp} using downturn LGD rates, as reported by the major Australian banks.  We would reach broadly the same conclusions using point-in-time LGD rates.

\begin{figure}[!t]
	\centering
	\caption[Impact of the financial crisis of 2007--09 on the Australian banking sector based on the allocation of credit losses between banking book exposures]{Impact of the financial crisis of 2007--09 on the credit portfolios of the major Australian banks.  Realisation $y$ of systematic risk factor~$Y$ describes the prevailing state of the economy.  Confidence level~$\alpha$ is the probability of the state of the economy being better than the economic scenario described by realisation~$y$.\protect\\
	(\textit{a}) Credit losses are allocated between IRB credit exposures and other banking book exposures in proportion to RWA.\protect\\
	(\textit{b}) Credit losses are allocated entirely to IRB credit exposures providing a lower bound on~$Y$.}
	\label{fig_y_alpha_rwa}
	\scalebox{0.90}{
	\begin{tabular}{c}
		\input{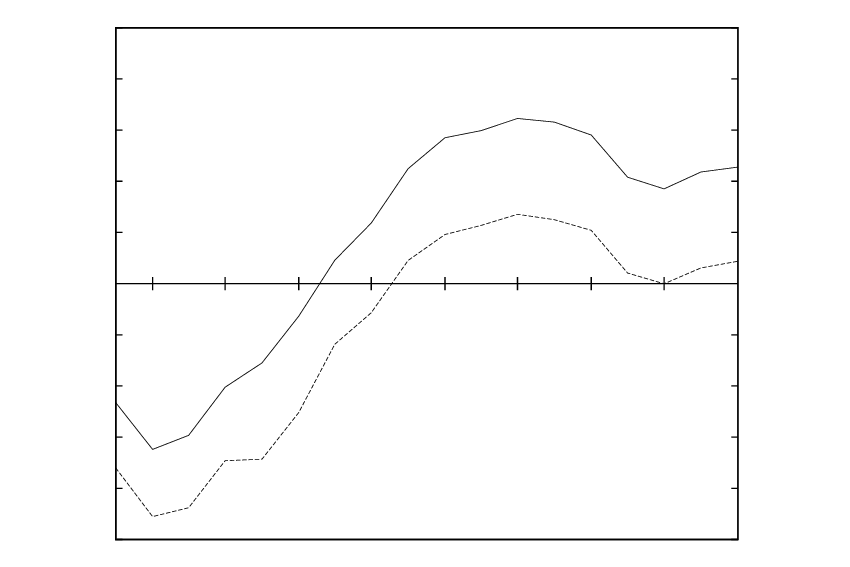}
	\end{tabular}
	}
\end{figure}

Real GDP growth, plotted for the Australian economy in Figure~\ref{fig_macro_econ}, is conventionally reported as seasonally adjusted, quarter-over-quarter.  Quarterly realisations of the single systematic risk factor are recovered by equating credit losses projected over a one-year horizon with credit losses incurred during the same one-year interval.  Accordingly, we restate quarterly observations of real GDP growth for the Australian economy as year-over-year.  Figure~\ref{fig_y_gdp} plots time series of the single systematic risk factor describing the prevailing state of the Australian economy and real GDP growth, year-over-year.  The time series are quite strongly correlated (+0.60), suggesting that the single systematic risk factor serves as a reasonable proxy for the relative state of the economy.  Indeed the correlation between the time series rises to +0.72 when realisations of the single systematic risk factor are lagged by one or two quarters.  Arguably, realisation~$y(t)$ of systematic risk factor~$Y$ leads real GDP growth, but realisation~$y(t)$ cannot be computed until credit losses $ \mathcal{L}_{r}(t\!+\!2)$ are recognised in profit and loss for quarter~$t\!+\!2$.  Overall, observations of the prevailing state of the Australian economy recovered from \eqref{eqn_cred_loss_cond_exp}, and plotted in Figures~\ref{fig_y_alpha_lgd} and \ref{fig_y_gdp}, agree rather well with the macroeconomic indicators and financial statistics plotted in Figure~\ref{fig_macro_econ}.

As noted above, we do not observe losses on IRB credit exposures.  Hence, we choose to allocate total credit losses incurred between IRB credit exposures and other banking book exposures in proportion to RWA.  Figure~\ref{fig_y_alpha_rwa} illustrates that even if we were to attribute credit losses entirely to IRB credit exposures and use downturn LGD rates in \eqref{eqn_cred_loss_cond_exp}, the financial crisis of 2007--09 would have inflicted credit losses on the Australian banking system not exceeding those incurred, on average, once every eight years --- a lower bound on the severity of the crisis in Australia.

\begin{figure}[!t]
	\centering
	\caption[Impact of the financial crisis of 2007--09 on the Australian banking sector taking into account delays in the recognition of bad debts]{Impact of the financial crisis of 2007--09 on the credit portfolios of the major Australian banks.  Sensitivity of the single systematic risk factor describing the prevailing state of the economy to delays in the recognition of bad debts.  Confidence level~$\alpha$ is the probability of the state of the economy being better than the economic scenario described by realisation~$y$ of systematic risk factor~$Y$.}
	\label{fig_y_alpha_lag}
	\scalebox{0.90}{
	\begin{tabular}{c}
		\input{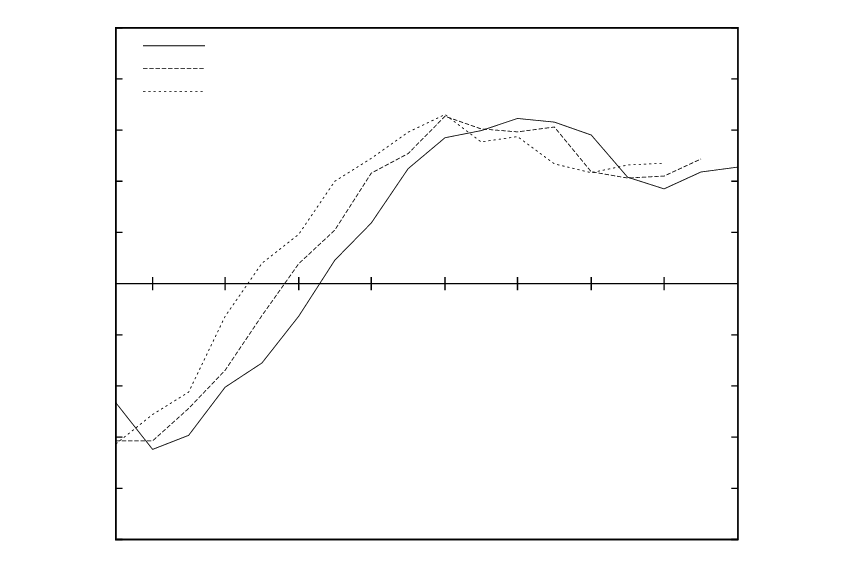}
	\end{tabular}
	}
\end{figure}

The model specification of the IRB approach was developed for the purpose of solvency assessment, or capital adequacy.  A credit risk model of capital adequacy requires precision in the measurement of absolute risk levels under stressed economic conditions associated with the tail of the portfolio loss distribution \citeyearpar[BCBS et al.,][]{JF25}.  Recognising that realisations of the single systematic risk factor describing states of the Australian economy experienced since the implementation of Basel~II correspond to observations away from the tail of its distribution, we cannot attest to the validity of the ASRF model for regulatory capital modelling.  However, we argue that our findings support a favourable assessment of the ASRF model, and asset value factor models of credit risk in general, for the purposes of capital allocation, performance attribution and risk monitoring.  These management functions, generally served by economic capital models, need only be accurate in the measurement of relative risk under ``normal'' economic conditions.

It could be argued that delays in the recognition of bad debts warrant introducing a lag in the association of projected credit losses with credit losses incurred.  Figure~\ref{fig_y_alpha_lag} plots the time series of the prevailing state of the economy (i.e., realisations of the single systematic risk factor) assuming no delay in the recognition of bad debts ($s = t$), along with time series assuming delays of one quarter ($s = t+1$) and two quarters ($s = t+2$).  Introducing lags to account for possible delays in the recognition of bad debts does not materially alter our measurement of the impact of the financial crisis on the credit portfolios of the major Australian banks at the depths of the crisis --- it remains, informally speaking, a one-in-five-year event (using downturn LGD rates).

\begin{figure}[!t]
	\centering
	\caption[Dating the inception of the financial crisis of 2007-09]{Dating the inception of the financial crisis of 2007--09.  Credit losses incurred by Australian banks peaked, and global equity indices plumbed their lows, after the implementation of Basel~II.\protect\\
		{\footnotesize Sources: Australian Prudential Regulation Authority; Bloomberg.}}
	\label{fig_gfc_incept}
	\scalebox{0.51}{
	\begin{tabular}{cc}
		\input{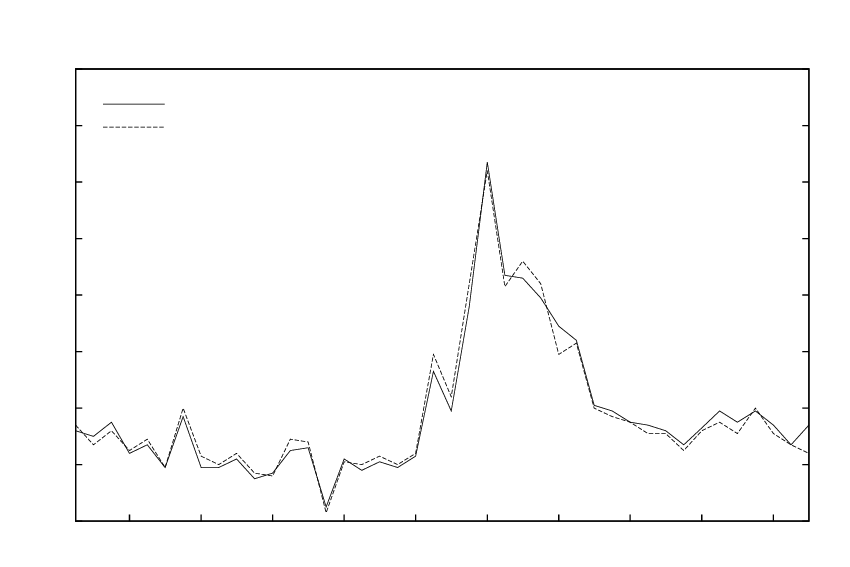}		& \input{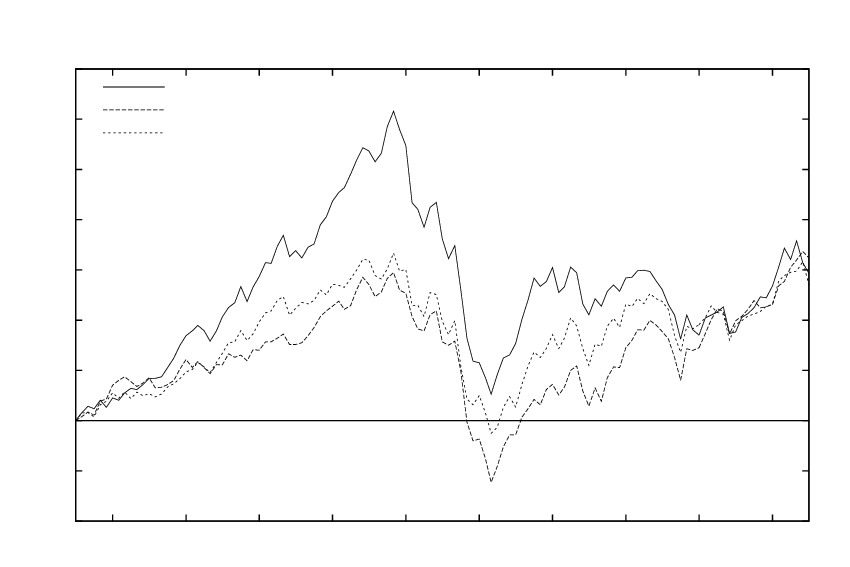} 
	\end{tabular}
	}
\end{figure}

This empirical analysis begins with the initial submission of capital adequacy and IRB credit risk forms as at 31~March 2008 and credit losses recognised in profit and loss between 1~April 2008 and 31~March 2009, and solves for the single systematic risk factor describing the state of the economy as at 30~September 2008.  Our measurement of the impact of the financial crisis on the Australian banking sector is necessarily predicated on the assumption that the effect of the crisis was most acutely felt after 31~March 2008.  Figure~\ref{fig_gfc_incept} indicates that credit losses incurred by the major Australian banks peaked in the fourth quarter of 2008 and remained elevated through 2009, and global equity indices plumbed their lows during the first quarter of 2009 --- S\&P ASX 200 index fell 54\% between November 2007 and March 2009.

Our measurement of the impact of the financial crisis on the Australian banking sector is buttressed by the macroeconomic stress test administered by APRA on the major Australian banks in 2012.  The three-year economic scenario developed for the stress test by APRA in conjunction with the RBA and Reserve Bank of New Zealand, was designed to be comparable with the actual experience of the United States and United Kingdom during the recent crisis.  In particular, the macroeconomic stress test envisaged: real GDP contracting 5\%; unemployment rising to 12\%; house prices falling 35\%; and commercial property prices falling 40\%.  None of the major Australian banks would have failed under this severe but plausible economic scenario, nor would any of the major banks have breached the 4\% minimum tier~1 capital requirement of the Basel~II Accord \citep{LJF12}.

\subsection{Australian Banks' Capacity to Absorb Credit Losses}\label{sect_dd}
Distance-to-default, which measures the level of capitalisation, reflects the capacity to absorb credit losses.  Substituting provisions set aside for absorbing expected losses and capital held against unexpected losses on IRB credit exposures into the ASRF model, and translating realisations of the single systematic risk factor into distance-to-default, we measure the level of capitalisation of the major Australian banks, in aggregate,\footnote{
Results are reported for the major Australian banks, in aggregate, so as not to violate confidentiality agreements.
} since the implementation of Basel~II.  

Suppose that banks set aside provisions and hold capital that are sufficient to absorb credit losses at the $\alpha$ confidence level.  Denote by $\mathcal{Q}_{r}(t)$ provisions set aside for absorbing expected credit losses as at the end of quarter~$t$, and $\mathcal{K}(t)$ the capital base as at the end of quarter~$t$, both reported on the capital adequacy form.  Recall that APRA sets a prudential capital ratio, subject to a minimum 8\% of RWA, for each ADI, and an ADI typically holds a capital buffer above its prudential capital requirement.  Since we observe neither provisions set aside for absorbing expected losses nor capital held against unexpected losses on IRB credit exposures, we choose an allocation procedure.  Provisions are allocated between IRB credit exposures and other banking book exposures in proportion to expected losses over the subsequent four quarters as projected at the end of quarter~$t$:
\begin{equation}\label{eqn_irb_prov}
	\mathcal{Q}_{n}(t) = \frac{\E\big[\mathcal{L}_{n}(t\!+\!1) + \mathcal{L}_{n}(t\!+\!2) + \mathcal{L}_{n}(t\!+\!3) + \mathcal{L}_{n}(t\!+\!4)\big]}{\E\big[\mathcal{L}_{r}(t\!+\!1) + \mathcal{L}_{r}(t\!+\!2) + \mathcal{L}_{r}(t\!+\!3) + \mathcal{L}_{r}(t\!+\!4)\big]}\mathcal{Q}_{r}(t).
\end{equation}
Capital is allocated between IRB credit exposures and other risk exposures (i.e., non-IRB credit exposures, and operational, market and securitisation risks) in proportion to RWA as at the end of quarter~$t$:
\begin{equation}\label{eqn_irb_cap}
	\mathcal{K}_{n}(t) = \frac{\mathcal{R}_{n}(t)}{\mathcal{R}(t)}\mathcal{K}(t).
\end{equation}
Hence, provisions set aside for absorbing expected losses and capital held against unexpected losses on IRB credit exposures as at the end of quarter~$t$ is equal to ${\mathcal{Q}_{n}(t) + \mathcal{K}_{n}(t)}$.

We estimate distance-to-default by solving for realisation~$\tilde{y}(t)$ of systematic risk factor~$Y$ that equates the sum of \eqref{eqn_irb_prov} and \eqref{eqn_irb_cap} to the conditional expectation of portfolio credit losses:\footnote{
We use the GNU Scientific Library routine \texttt{gsl\_root\_fsolver\_falsepos}, which implements the \textit{false position algorithm}, to solve for $\tilde{y}(t)$.
}
\begin{equation}\label{eqn_solver_dd}
	\mathcal{Q}_{n}(t) + \mathcal{K}_{n}(t) = \sum_{i=1}^{n} \ead_{i}(t)\lgd_{i}(t)\nu_{i}(t) \Phi\left(\frac{\Phi^{-1}\big(p_{i}(t)\big) - \sqrts{\rho_{i}(t)}\tilde{y}(t)}{\sqrts{1-\rho_{i}(t)}}\right).
\end{equation}
Recall that APRA's prudential standards instruct ADIs to assign unconditional PDs and downturn LGD rates to IRB credit exposures.  Since distance-to-default reflects the capacity to absorb credit losses, usually under stressed economic conditions, it is appropriate to use downturn LGD rates to estimate distance-to-default.  Hence, $\lgd_{i}(t) = \underline{\lgd_{i}}(t)$ for $i = 1, \dots, n$ and for all $t$ in \eqref{eqn_solver_dd}.  Note that the sum of provisions set aside for absorbing expected credit losses and capital held against unexpected credit losses in~\eqref{eqn_solver_dd} replaces the sum of expected losses and regulatory capital in the ASRF model described in Section~\ref{sect_model_spec_irb}.  Denote by $\tilde{d}(t)$ the distance-to-default as at the end of quarter~$t$.  Then,
\begin{equation}\label{eqn_dd}
	\tilde{d}(t) = -\tilde{y}(t),
\end{equation}
which translates into confidence level
\begin{equation}\label{eqn_alpha_dd}
	\tilde{\alpha}(t) = 1 - \Phi\big(-\tilde{d}(t)\big) = \Phi\big(\tilde{d}(t)\big).
\end{equation}
It follows from~\eqref{eqn_irb_cap} that the risk-based capital ratio as at the end of quarter~$t$, plotted in Figure~\ref{fig_cap_ratio}, is given by
\begin{equation}\label{eqn_cap_ratio}
	\kappa(t) = \frac{\mathcal{K}(t)}{\mathcal{R}(t)} =  \frac{\mathcal{K}_{n}(t)}{\mathcal{R}_{n}(t)}.
\end{equation}

\afterpage{
\begin{landscape}
\begin{table}[!t]
	\centering
	\caption[Risk-based capital ratio and distance-to-default of the major Australian banks]{As reported risk-based capital ratio~$\kappa(t)$, and implied distance-to-default~$\tilde{d}(t)$ of the major Australian banks, in aggregate.  The reverse stress test uncovers the weakest economic shock~$\hat{y}(t)$ that would result in a breach of the capital ratio floor~$\underline{\kappa}$.  Realisation~$y(t)$ describes the prevailing state of the economy.}
	\label{tbl_cap_ratio_dd}
	\rowcolors{2}{Gold}{}
	\begin{tabular}[c]{c r r r c r r c r r c r r}
		\hiderowcolors
		\toprule
		& \multicolumn{3}{c}{}	&	& \multicolumn{5}{c}{Reverse stress test}	&	& \multicolumn{2}{c}{\multirow{2}{2.50cm}[-2.5pt]{Prevailing state of the economy}}	\\
		\cmidrule{6-10}	
		& \multicolumn{3}{c}{As reported}	&	& \multicolumn{2}{c}{$\underline{\kappa} = 4.0\%$}	&	& \multicolumn{2}{c}{$\underline{\kappa} = 8.0\%$}	&	& \multicolumn{2}{c}{} 	\\
		\cmidrule{2-4}\cmidrule{6-7}\cmidrule{9-10}\cmidrule{12-13}
		 \multicolumn{1}{c}{$t$} 	& \multicolumn{1}{c}{$\kappa(t)$, \%}		& \multicolumn{1}{c}{$\tilde{d}(t)$}	& \multicolumn{1}{c}{$\tilde{\alpha}(t)$, \%}	& 	& \multicolumn{1}{c}{$\hat{y}(t)$}	&  \multicolumn{1}{c}{$\hat{\alpha}(t)$, \%}	&	& \multicolumn{1}{c}{$\hat{y}(t)$}	& \multicolumn{1}{c}{$\hat{\alpha}(t)$, \%}		&	&  \multicolumn{1}{c}{$y(t)$}	&  \multicolumn{1}{c}{$\alpha(t)$, \%}	\\
		\midrule
		\showrowcolors
		31-Mar-2008 	& 10.67	& 3.504	& 99.977	&	& $-2.915$	& 99.822	&	& $-1.90$	& 97.10	&	&		&		\\
		30-Jun-2008	& 10.73	& 3.413	& 99.968	&	& $-2.845$	& 99.778	&	& $-1.87$	& 96.94	&	&		&		\\
		30-Sep-2008 	& 10.98	& 3.508	& 99.978	&	& $-2.946$	& 99.839	&	& $-2.01$	& 97.78	& 	&$-0.58$	& 71.9	\\
		31-Dec-2008 	& 11.51	& 3.588	& 99.983	&	& $-3.044$	& 99.883	&	& $-2.17$	& 98.49	&	& $-0.81$	& 79.1	\\
		31-Mar-2009 	& 11.38	& 3.621	& 99.985	&	& $-3.044$	& 99.883	&	& $-2.12$	& 98.29	&	& $-0.74$	& 77.1	\\
		30-Jun-2009	& 11.10	& 3.620	& 99.985	&	& $-3.018$	& 99.873	& 	& $-2.04$	& 97.94	&	& $-0.51$ & 69.4	\\
		30-Sep-2009 	& 11.52	& 3.682	& 99.988	&	& $-3.095$	& 99.902	&	& $-2.17$	& 98.51	&	& $-0.39$	& 65.1	\\
		31-Dec-2009 	& 11.85	& 3.732	& 99.991	&	& $-3.160$	& 99.921	& 	& $-2.28$	& 98.88	&	& $-0.16$	& 56.3	\\
		31-Mar-2010 	& 11.82	& 3.717	& 99.990	&	& $-3.145$	& 99.917	&	& $-2.26$	& 98.82	&	& 0.11	& 45.5	\\\
		30-Jun-2010	& 11.51	& 3.677	& 99.988	&	& $-3.090$	& 99.900	&	& $-2.17$	& 98.48	&	& 0.30	& 38.4	\\
		30-Sep-2010 	& 11.50	& 3.680	& 99.988	&	& $-3.092$	& 99.901	&	& $-2.16$	& 98.48	&	& 0.56	& 28.7	\\
		31-Dec-2010 	& 11.27	& 3.653	& 99.987	&	& $-3.057$	& 99.888	&	& $-2.10$	& 98.20	&	& 0.71	& 23.8	\\
		31-Mar-2011 	& 11.49	& 3.682	& 99.989	&	& $-3.098$	& 99.902	&	& $-2.17$	& 98.49	&	& 0.75	& 22.7	\\
		30-Jun-2011	& 11.42	& 3.660	& 99.987	&	& $-3.076$	& 99.895	&	& $-2.14$	& 98.38	&	& 0.81	& 21.0	\\
		30-Sep-2011 	& 11.48	& 3.647	& 99.987	&	& $-3.083$	& 99.897	&	& $-2.18$	& 98.54	&	& 0.79	& 21.5	\\
		31-Dec-2011 	& 11.42	& 3.642	& 99.986	&	& $-3.077$	& 99.895	&	& $-2.17$	& 98.51	&	& 0.73	& 23.4	\\
		31-Mar-2012 	& 11.48	& 3.637	& 99.986	&	& $-3.077$	& 99.896	&	& $-2.18$	& 98.55	&	& 0.52	& 30.2	\\
		30-Jun-2012	& 11.44	& 3.633	& 99.986	&	& $-3.075$	& 99.895	&	& $-2.18$	& 98.54	&	& 0.46	& 32.2	\\
		30-Sep-2012 	& 11.72	& 3.659	& 99.987	&	& $-3.121$	& 99.910	&	& $-2.28$	& 98.86	&	& 0.54	& 29.3	\\
		31-Dec-2012 	& 11.75	& 3.663	& 99.988	&	& $-3.126$	& 99.911	&	& $-2.29$	& 98.89	&	& 0.57	& 28.5	\\
		31-Mar-2013 	& 11.69	& 3.698	& 99.989	&	& $-3.165$	& 99.923	&	& $-2.34$	& 99.04	&	&		&		\\
		30-Jun-2013	& 11.42	& 3.657	& 99.987	&	& $-3.118$	& 99.909	&	& $-2.27$	& 98.83	&	&		&		\\
		\bottomrule
	\end{tabular}
\end{table}
\end{landscape}
}

Table~\ref{tbl_cap_ratio_dd} reports quarterly risk-based capital ratio of the major banks, in aggregate, along with implied distance-to-default.  Since 2008 they have maintained a capital base that is consistent with targeting a credit rating between A and AA (i.e., a target confidence level between $99.96\%$ and $99.99\%$).  Note that, under the prudential standards of APRA \citeyearpar{APS113}, if provisions set aside are insufficient to absorb expected credit losses, the shortfall is deducted from capital.  The general agreement of our estimates of distance-to-default with credit ratings issued by external rating agencies seemingly lends further support to a favourable assessment of the model specification of the IRB approach.  However, distance-to-default is a measure of capital adequacy, and Section~\ref{sect_y} argues that we cannot attest to the suitability of the ASRF model for the purpose of solvency assessment, because realisations of the single systematic risk factor experienced in Australia since the implementation of Basel~II correspond to observations away from the tail of its distribution.  Therefore, we report estimates of distance-to-default with the caveat that \eqref{eqn_solver_dd} models default dependence in the tail of the portfolio loss distribution as a multivariate Gaussian process.  It is generally acknowledged that models which assume that financial data follow a Gaussian distribution tend to underestimate tail risk.  If we were to model default dependence by heavier-tailed distributions, we would recover narrower estimates of distance-to-default.  Also, credit losses incurred at the depths of the financial crisis would correspond to a quantile of the portfolio loss distribution further from the tail, and thus associated with a less contractionary, or more expansionary, state of the economy.

In contrast to our evaluation of the capacity of the major banks to absorb credit losses during the financial crisis, \citet{AP12} argue that Australian banks did not fare much better than their global counterparts.  They compared the distance-to-default of the Australian banking sector with that of the banking sectors in the United States, Europe and Canada.  In their implementation of the KMV/Merton structural methodology, distance-to-default is a function of implied market value and volatility of assets imputed from equity prices.  \citeauthor{AP12} observed that the distance-to-default for Australian banks narrowed sharply from a peak of 11.31 in 2005 to 0.60 in 2008.  Note that a distance-to-default of $0.60$ translates into roughly a one-in-four chance of bank failure, but in reality there were none.  An implicit assumption of their analysis is that security markets consistently price risk fairly.  But it's not uncommon for markets to exhibit bouts of manic-depressive behaviour.  During periods when market participants are gripped by fear (respectively, driven by greed), their perception of risk is heightened (lessened), and markets become undervalued (overpriced).  We submit that their results are biased by plummeting equity prices and spiking volatility reflecting the overreaction of market participants gripped by fear at the depths of the crisis --- the S\&P ASX 200 Banks index fell 58\% between November 2007 and January 2009.  With access to internal bank data collected by APRA, we produce a fundamental evaluation of the effect of the recent crisis and solvency of Australian banks that differs markedly from the stock market's assessment.  

\subsection{Reverse Stress Testing}\label{sect_rev_stress_test}
Stress testing is an important risk management tool promoted by supervisors through the Basel~II framework.  In its principles for sound stress testing, BCBS \citeyearpar{BCBS155} recommends that supervisors make regular and comprehensive assessments of banks' stress testing programs, and encourages supervisors to conduct stress tests based on common scenarios for banks in their jurisdiction.

A key principle of APRA's supervisory review is that regulated institutions are responsible for developing and maintaining an internal capital adequacy assessment process (ICAAP) proportional to the size, business mix and complexity of their operations.  ICAAP is an integrated and documented approach to risk and capital management that assesses the risk appetite of an institution and establishes the commensurate level and quality of capital.  As part of ICAAP, institutions are expected to conduct their own and supervisory-led stress tests.  APRA believes that stress testing goes beyond assessing capital adequacy to informing risk appetite, setting risk limits, identifying vulnerabilities and developing mitigating actions.  The results of stress tests are used by supervisors ``to anchor expectations for the level of capital that an institution should hold in normal times to provide a sufficient buffer to withstand a challenging environment.''  The results also serve ``to inform our [supervisors'] risk assessment of institutions as part of the development of supervisory action plans'' \citep{LJF12}.

Usually, a stress test begins with the development of an economic scenario that is the manifestation of some stress event or economic shock.  It is then translated into conditional (point-in-time) PDs assigned to obligors constituting a bank's credit portfolio.  Finally, an analytical or simulation model estimates portfolio credit losses subject to the stress event, which are charged against provisions and capital to produce an assessment of the bank's solvency.  A variation on this methodology is reverse stress testing --- a technique in which ``losses that would render an institution unviable or subject to material regulatory intervention are identified and attention then focussed on the types of scenarios that would generate these losses'' \citep{LJF10}.

Clearly, realisation~$\tilde{y}(t) = -\tilde{d}(t)$, recovered from~\eqref{eqn_solver_dd}, describes an economic scenario that would render the major banks, in aggregate, insolvent at the end of quarter~$t$ if the associated credit losses were incurred and recognised instantaneously.  We extend this rudimentary form of reverse stress testing to uncover economic scenarios that would cause the major banks to breach some designated capital ratio floor~$\underline{\kappa}$, and consequently trigger material supervisory intervention.  Suppose that the expectation of portfolio credit losses conditional on stress event~$\hat{y}(t)$ is instantaneously incurred and recognised in profit and loss at the end of quarter~$t$.  Then, the capital ratio floor would be breached at the end of quarter~$t$ if credit losses exceeded
\begin{equation}\label{eqn_loss_cap_ratio_floor}
	\mathcal{Q}_{n}(t) + \mathcal{K}_{n}(t) - \underline{\kappa}\mathcal{R}_{n}(t).
\end{equation}
Economic scenarios worse than that described by realisation~$\hat{y}(t)$ of systematic risk factor~$Y$ satisfying~\eqref{eqn_solver_cap_ratio_floor} would result in a breach of the capital ratio floor at the end of quarter~$t$:\footnote{
We use the GNU Scientific Library routine \texttt{gsl\_root\_fsolver\_falsepos}, which implements the \textit{false position algorithm}, to solve for $\hat{y}(t)$.
}
\begin{equation}\label{eqn_solver_cap_ratio_floor}
	\mathcal{Q}_{n}(t) + \mathcal{K}_{n}(t) - \underline{\kappa}\mathcal{R}_{n}(t) = \sum_{i=1}^{n} \ead_{i}(t)\lgd_{i}(t)\nu_{i}(t) \Phi\left(\frac{\Phi^{-1}\big(p_{i}(t)\big) - \sqrts{\rho_{i}(t)}\hat{y}(t)}{\sqrts{1-\rho_{i}(t)}}\right).
\end{equation}
Then, the probability of the state of the economy being better than the economic scenario described by realisation~$\hat{y}(t)$ is a most
\begin{equation}\label{eqn_alpha_cap_ratio_floor}
	\hat\alpha(t) = 1 - \Phi\big(\hat{y}(t)\big).
\end{equation}
Finally, macroeconomic-based models would translate realisations of the single systematic risk factor into observations of macroeconomic indicators (e.g., real GDP growth, unemployment rate, house prices, etc.).  This final translation is beyond the scope of this paper.

Our rudimentary stress testing methodology has its limitations: it is static; focusses exclusively on credit risk, assuming away market, liquidity and operational risks; fails to consider diversification benefits or compounding effects arising from the interaction between risk classes; excludes net interest income earned during the period in which credit losses are incurred; and does not consider the possibility of raising fresh capital.  Yet it puts the severity of the financial crisis, as experienced by the Australian banking sector, in perspective by comparing it with stress events that would trigger material supervisory intervention.  

Table~\ref{tbl_cap_ratio_dd} reports the weakest economic shock~$\hat{y}(t)$ imparted at the end of quarter~$t$ that would result in a breach of capital ratio floors set at $4.0\%$ and $8.0\%$ if credit losses were instantaneously incurred and recognised in profit and loss.  Section~\ref{sect_y} made reference to the macroeconomic stress test administered by APRA on the major banks in 2012, describing the severe and plausible economic scenario under which none of the major banks would have breached the 4\% minimum tier 1 capital requirement.  Table~\ref{tbl_cap_ratio_dd} indicates that with $\underline{\kappa} = 4.0\%$, approximately $99.9\%$ of economic scenarios, presumably including the severe but plausible one developed for the macroeconomic stress test administer by APRA, would not have resulted in the major banks, in aggregate, breaching the capital ratio floor in 2012.  Even with $\underline{\kappa} = 8.0\%$, fewer than $1.5\%$ of economic scenarios would have resulted a breach of the capital ratio floor in 2012.  During the recent crisis the capacity of the major banks, in aggregate, to absorb credit losses appears to have been ``stretched'' to the point where $0.22\%$ (respectively, $3.06\%$) of economic scenarios would have resulted in a breach of the $4.0\%$ ($8.0\%$) capital ratio floor during the quarter ending 30~June 2008.  For comparison, Table~\ref{tbl_cap_ratio_dd} also reports the prevailing state of the economy described by quarterly realisations~$y(t)$ recovered from \eqref{eqn_cred_loss_cond_exp} using downturn LGD rates, and plotted in Figures~\ref{fig_y_alpha_lgd} and~\ref{fig_y_gdp}. 

\section{Conclusion}\label{sect_concl}
This paper examines the model specification of the Basel~II IRB approach.  Our empirical analysis takes measurements from the ASRF model of the Australian banking sector, and compares them with signals from macroeconomic indicators, financial statistics and external credit ratings to render a fundamental assessment.  We believe that our findings are applicable to the broader class of asset value factor models of credit risk.  While our fundamental assessment makes observations about the prevailing state of Australia's economy and the solvency of its banking sector, we do not propose model enhancements, nor draw policy implications.  Although, our observations may open a debate on the model specification of the IRB approach or related regulatory policies.  

The ASRF model calculates the expectation of credit losses conditional on a realisation of the single systematic risk factor, which is interpreted as describing the state of the economy, in order to assess regulatory capital charges.  Substituting credit losses incurred into the ASRF model, we solve for realisations of the single systematic risk factor generating a time series of the prevailing state of the Australian economy since the implementation of Basel~II.  Its moderately strong correlation with the time series of real GDP growth supports a favourable assessment of the ASRF model for the purposes of capital allocation, performance attribution and risk monitoring.  Since the depths of the financial crisis of 2007--09 were reached after the implementation of Basel~II, our empirical analysis reveals that the economic shock initiated by the crisis imparted a mild stress on the Australian banking sector.  Then, substituting provisions set aside for absorbing expected credit losses and capital held against unexpected credit losses into the ASRF model, and translating realisations of the single systematic risk factor into distance-to-default, we measure the level of capitalisation of the major Australian banks.  In aggregate, they have maintained a capital base that is reflective of the strong credit ratings issued by external rating agencies.  Reverse stress testing, too, reveals that the major banks were adequately capitalised to absorb a substantially more severe shock than the one imparted by the recent crisis.  Our fundamental assessment of the impact of the financial crisis on the credit portfolios of the major banks differs markedly from the stock market's assessment, which we argue is biased by the overreaction of market participants gripped by fear at the depths of the crisis. 

We reiterate that our estimates of the prevailing state of the economy and level of capitalisation of the major banks since the implementation of Basel~II are recovered from the ASRF model prescribed under the IRB approach, which models default dependence as a multivariate Gaussian process.  However, it is generally acknowledged that models which assume that financial data follow a Gaussian distribution tend to underestimate tail risk.  If we were to model default dependence by heavier-tailed distributions, we would find that credit losses incurred at the depths of the financial crisis correspond to a quantile of the portfolio loss distribution further from the tail, and thus associated with a less contractionary, or more expansionary, state of the economy.  Also, heavier-tailed distributions would imply a narrower distance-to-default.

Recognising that realisations of the single systematic risk factor describing states of the Australian economy experienced since the implementation of Basel~II correspond to observations away from the tail of its distribution, we cannot attest to the accuracy of the ASRF model in the tail of the portfolio loss distribution, nor its validity for regulatory capital modelling.  An evaluation of the ASRF model for the purpose of solvency assessment, would involve taking readings of north Atlantic banking jurisdictions that experienced the full force of the financial crisis of 2007--09.  It is unlikely to be as favourable as the qualified evaluation presented here.  The UK banking sector, which has operated under the Basel~II framework since 2008, experienced losses during the crisis on a scale that led to the failure or nationalisation of large banks including Bradford \& Bingley, HBOS, Lloyds Banking Group, Northern Rock and Royal Bank of Scotland.  Moreover, the Basel~2.5 and Basel~III reform packages have been developed by BCBS to address deficiencies of the Basel~II framework exposed by the recent crisis.

From our assessment of the model specification of the IRB approach emerges a methodology for regulators to monitor the prevailing state of the economy as described by the single systematic risk factor, and the capacity of supervised banks to absorb credit losses as measured by distance-to-default.  While confidentiality agreements preclude us from publishing results for individual banks, regulators would be at liberty to conduct their analysis on an individual bank basis.  Measurements from the ASRF model signalling an overheating economy and procyclical movements in capital bases, corroborated by macroeconomic performance indicators, would prompt supervisory intervention.  For example, banks could be instructed to build up their countercyclical capital buffer introduced under Basel~III in order to rein in rapidly accelerating credit growth.  Furthermore, a longer history of the time series of the prevailing state of the economy recovered from the ASRF model could serve to validate and calibrate macroeconomic-based models for estimating conditional (point-in-time) PDs.  Macroeconomic-based models for estimating point-in-time PDs usually express the single systematic risk factor as a function of macroeconomic variables and a random economic shock \citep{CLJA06}.  This time series could also inform the development of economic scenarios for stress testing, a standard tool of prudential supervision.

\section*{Acknowledgements}
This research has been conducted with the support of the Australian Prudential Regulation Authority (APRA).  It has provided access to professionals with expertise in financial regulation, risk modelling and data management, along with access to internal bank data collected by APRA from the institutions that it supervises.  We acknowledge with pleasure the support and encouragement from Charles Littrell, Executive General Manager at APRA, as well as his constructive comments on early drafts of this paper.  Anthony Coleman and Guy Eastwood of the Credit Risk Analytics team offered valuable input in the design of the empirical analysis and review of the findings.  The Statistics team, led by Steve Davies, provided data support for the empirical analysis, and reviewed the paper for compliance with confidentiality agreements.  Finally, we thank the Research team, headed by Bruce Arnold, for its contribution in distilling research topics that are pertinent to prudential regulators.   Financial support for this research is gratefully received in the form of an Australian Postgraduate Award, and a scholarship sponsored by the Capital Markets Cooperative Research Centre and APRA.



\end{document}